\documentclass[twocolumn,article,superscriptaddress,groupedaddress]{revtex4}
\usepackage[export]{adjustbox}
\usepackage{booktabs}
\usepackage{units}
\usepackage{graphicx}
\usepackage{enumitem}
\usepackage{xcolor}
\usepackage{hyperref}
\usepackage{psfrag}
\usepackage{graphicx}
\usepackage{dcolumn}
\usepackage{bm}
\usepackage{textcomp}

\begin{document}

\title{Cryogenic thermo-acoustic oscillations highlight and study in the SPIRAL2 superconducting LINAC}

\author{Adnan Ghribi}\email{ghribi@in2p3.fr}
\affiliation{Grand Acc\'el\'erateur National d'Ions Lourds (GANIL)}%
\affiliation{Centre National de la Recherche Scientifique (CNRS - IN2P3)}%
\author{Muhammad Aburas}%
\affiliation{Grand Acc\'el\'erateur National d'Ions Lourds (GANIL)}%
\affiliation{Commissariat à l'Energie Atomique (CEA - IRFU)}%
\author{Yoann Baumont}%
\affiliation{Grand Acc\'el\'erateur National d'Ions Lourds (GANIL)}%
\affiliation{Commissariat à l'Energie Atomique (CEA - IRFU)}%
\author{Pierre-Emmanuel Bernaudin}%
\affiliation{Grand Acc\'el\'erateur National d'Ions Lourds (GANIL)}%
\affiliation{Commissariat à l'Energie Atomique (CEA - IRFU)}%
\author{Stéphane Bonneau}%
\affiliation{Grand Acc\'el\'erateur National d'Ions Lourds (GANIL)}%
\affiliation{Centre National de la Recherche Scientifique (CNRS - IN2P3)}%
\author{Jean-François Leyge}%
\author{Guillaume Lescalié}%
\affiliation{Grand Acc\'el\'erateur National d'Ions Lourds (GANIL)}%
\affiliation{Commissariat à l'Energie Atomique (CEA - IRFU)}%
\author{Jean-Pierre Thermeau}%
\affiliation{Laboratoire AstroParticule et Cosmologie}%
\affiliation{Centre National de la Recherche Scientifique (CNRS - IN2P3)}%
\author{Yann Thivel}%
\affiliation{Grand Acc\'el\'erateur National d'Ions Lourds (GANIL)}%
\affiliation{Commissariat à l'Energie Atomique (CEA - IRFU)}%
\author{Laurent Valentin}%
\affiliation{Grand Acc\'el\'erateur National d'Ions Lourds (GANIL)}%
\affiliation{Centre National de la Recherche Scientifique (CNRS - IN2P3)}%
\author{Adrien Vassal}%
\affiliation{Grand Acc\'el\'erateur National d'Ions Lourds (GANIL)}
\affiliation{Commissariat à l'Energie Atomique (CEA/INAC-SBT)}%
\affiliation{Univ. Grenoble Alpes}%
\affiliation{Univ. Caen Basse Normandie}

\date{\today}

\begin{abstract}
Cryogenic thermoacoustic oscillations is an area of interest of several studies. For superconducting accelerators, it is an unwanted phenomena that we usually want to get rid off. The SPIRAL2 superconducting accelerator had distributed Taconis all over its cryostats. This paper spans the different steps from their first detection to their damping with a highlight on the methods and the instrumentation that has been used. The presented study also sets the ground for a real life experimental investigation of thermo-ascoustics in complex geometries such as superconducting LINACs. With modern Big data analysis and simulation tools, it also sets the ground for the developments of new data linked codes that fit these complex situations.

\end{abstract}

\maketitle

\section{\label{sec1}Introduction}
SPIRAL2 is a new generation heavy ions accelerator. Its heart is a superconducting linear accelerator (LINAC). 19 cryostats (called cryomodules) are spread along the beam line and encompass the accelerating structures (superconducting quater wave resonating cavities) \cite{Lewitowicz:2006fx, ferdinand:in2p3-00867502, dolegieviez:hal-02187926}. They are connected with a cryodistribution line and fed with near atmospheric pressure liquid helium (~4 K) for the main cavities and 14 bars 60 K helium gas for the thermal screens. The SPIRAL2 cryoplant centred around an Air Liquide Helial LF cold box manages to furnish the necessary cooling power. It can manage 120 g/s and 1100 W at 4 K \cite{ghribi2017, ghribi2017_2}.  It is worth mentioning here that the cryomodules are of two kinds/families. 12 of them enclose a single cavity and are called type A. The other 7 enclose two cavities and are called type B. The valves boxes that manage the fluids of these cryomodules also have some geometrical differences.
One of the main roles of the cryogenic system is to maintain stable conditions compatible with the operation constraints. That is to maintain all the cavities at a stable and uniform temperature (plunged in liquid helium) and the pressure variation within ± 5 mbars. If the liquid helium level drops in a cryomodule liquid helium phase separator, there is a risk that the corresponding cavity quenches, ie. loses its superconducting state. If the pressure in the phase separators varies too much and too quickly, the efforts applied on the cavities surface result in an elastic deformation of its shape. That would change its impedance in a way that the cavity wouldn't be matched anymore to its frequency of operation. There are of course a number of corrections to impedance or RF phase changes. For instance, the Low Level Radio Frequency (LLRF) system feeding the RF power to the cavity can manage a certain bandwidth at the cost of some power margin. This correction is fast and limited to small variations. The frequency tuning system can manage another correction, this time slow (more than one second) and adapted to large variations within its range of operation. The third and final way to limit the impedance variations of the cavities is simply by controlling the pressure in the phase separators. This stringent requirement for a biphasic cryogenic operation has led to several model based developments of the way input and output valves can be controlled \cite{Vassal_2019, Bonne_2020}.

When we first fed RF power to the cavities, we noticed a strange behaviour in the amplitude and phase of both transmitted and reflected RF power. After deep investigations, it turned out that what we were measuring was due to cryogenic thermo-acoustic oscillations (TAO) spread all over the LINAC.

This phenomenon named after K. W. Taconis who first encountered it in 1945 \cite{TACONIS1949733} is know to occur in a cryogenic system for specific geometry, temperature and pressure conditions \cite{Swift2007}. They have been thoroughly studied \cite{kramers1949, rott1969, rott1973, doi:10.1063/1.2990763, SUN201638} but remain difficult to predict and to deal with.

This paper relates the investigation, discovery and suppression of these oscillations with a highlight on the methods and instrumentation used. In the first section, we introduce the first discovery of the TAO on the SPIRAL2 cryogenic system. We then describe the experimental methods used to determine its frequency, amplitude and conditions of occurence. Finally we describe the trials and experimental solutions tested, the final chosen correction, the surveillance system that we put in place with its future evolution and numerical and theoretical considerations on the subject.

\section{First encounters}
\begin{figure}
  \includegraphics[width=.5\textwidth, right]{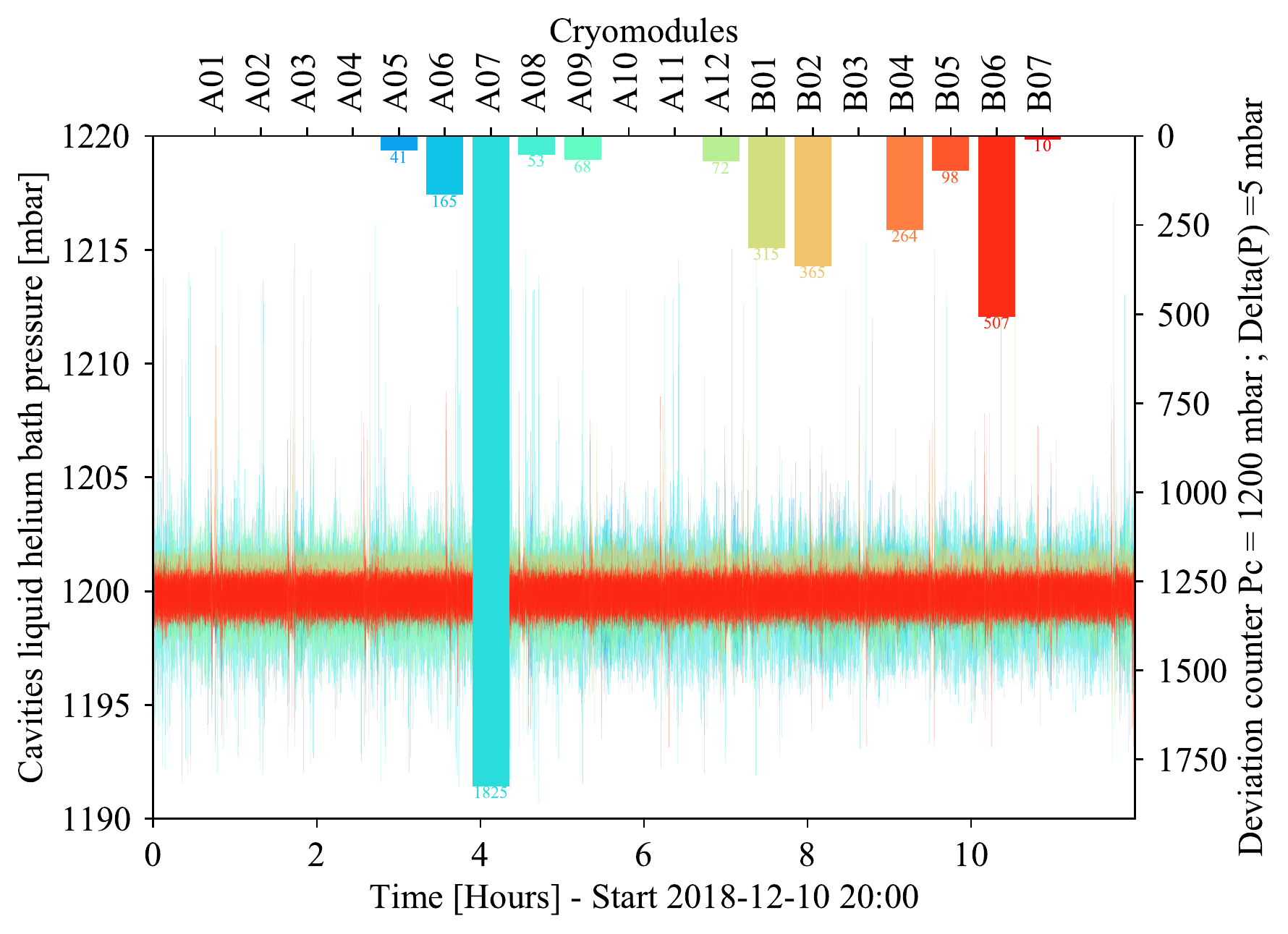}
  \caption{Top : pressure deviation counter with a threshold set to $\pm 5$ mbars. Bottom : absolute pressure measurements of the liquid helium phase separators of the LINAC superconducting cavities. The different colors are a signature of the cavities/cryomodules positions in the LINAC.}\label{pt_001_2018}
\end{figure}

The first TAO that we discovered was in the main 5000~L liquid helium dewar of the cryoplant in 2016. These oscillations were hard to miss : high thermal load, ice on the dewar neck, sometimes hearable continuous noise and the fast displacement of the pointer needle of the pressure gauge. To better estimate and quantify the phenomenon, we had to bypass the cold box Programmable Logic Controller (PLC) acquisition system that was too slow to measure the occurring phenomenon. We tried several approaches to suppress the dewar TAO : investigation of all the room temperature output ports with sometimes long lines used for filling or supplying liquid helium (LHe) from or to small mobile dewars, short circuit between the vapour sky and the room temperature lines and buffer capacities of several dimensions. Finally the approach that worked the best was the use of a 5 L buffer capacity in parallel with a micrometric valve \cite{ghribi:in2p3-01569768}. This TAO was at 100 Hz and could not be seen with regular slow absolute pressure transmitters. However, its effect could be seen also on cavities pressure. Even after correction of this TAO, some behaviour in the LINAC remained unexplained. This took the form of pressure glitches and pressure instabilities that propagated to all cavities with no prior notice. This behaviour made the pressure regulation very difficult, as if the absolute pressure sensors used by the valves to control the pressure was not giving a reliable information. Figure \ref{pt_001_2018} shows an example of such behaviour.

\begin{figure}[hbt]
  \includegraphics[width=.5\textwidth, right,trim={0cm 0cm 0cm 1cm},clip]{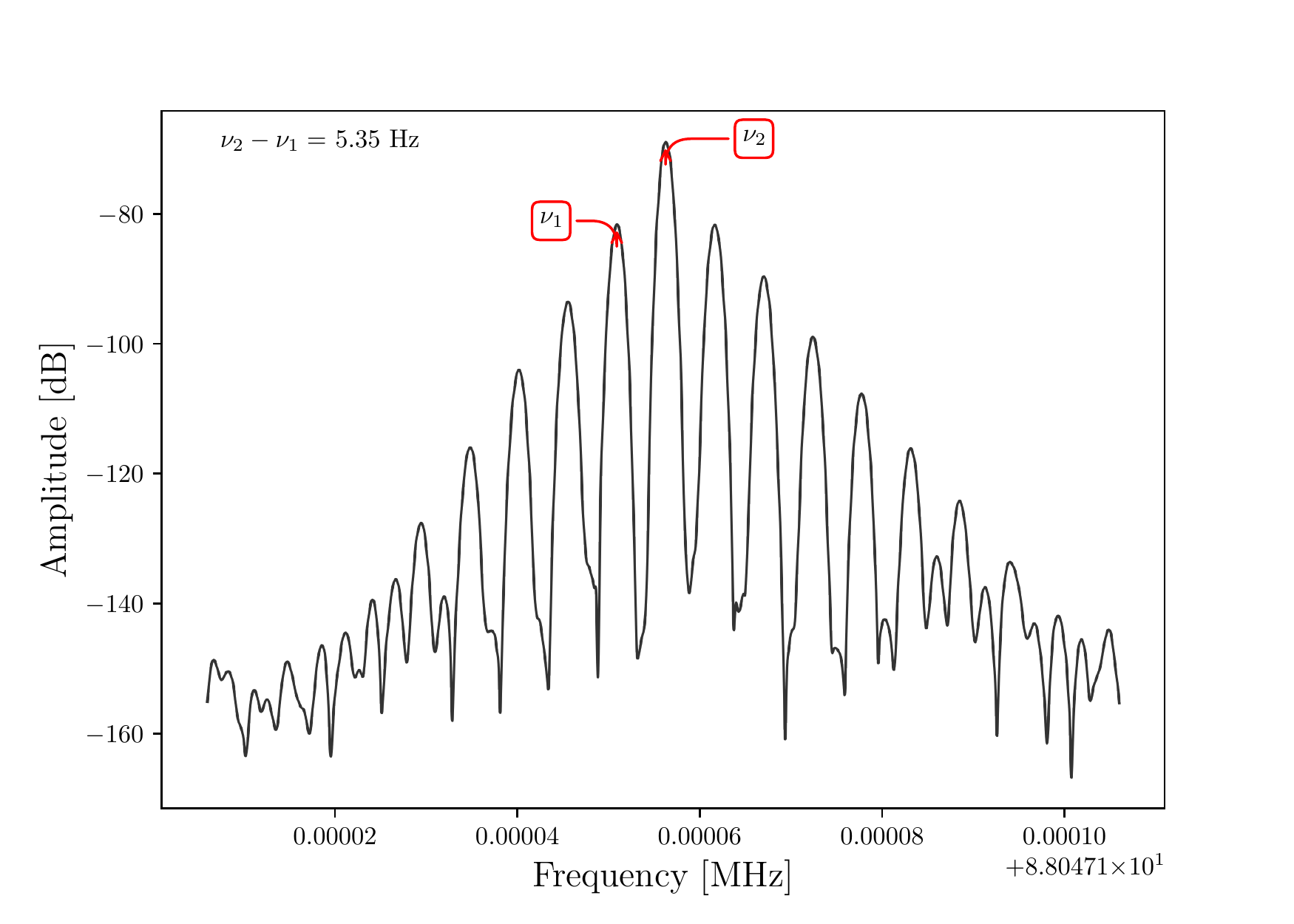}
  \caption{Amplitude modulation of the input/output signal difference for cryomodule CMB07 as reported in \cite{GHRIBI2020103126}.}\label{spectre_rf}
\end{figure}

\begin{figure}[hbt]
  \includegraphics[width=.48\textwidth, right,trim={4cm 6cm 5cm 7cm},clip]{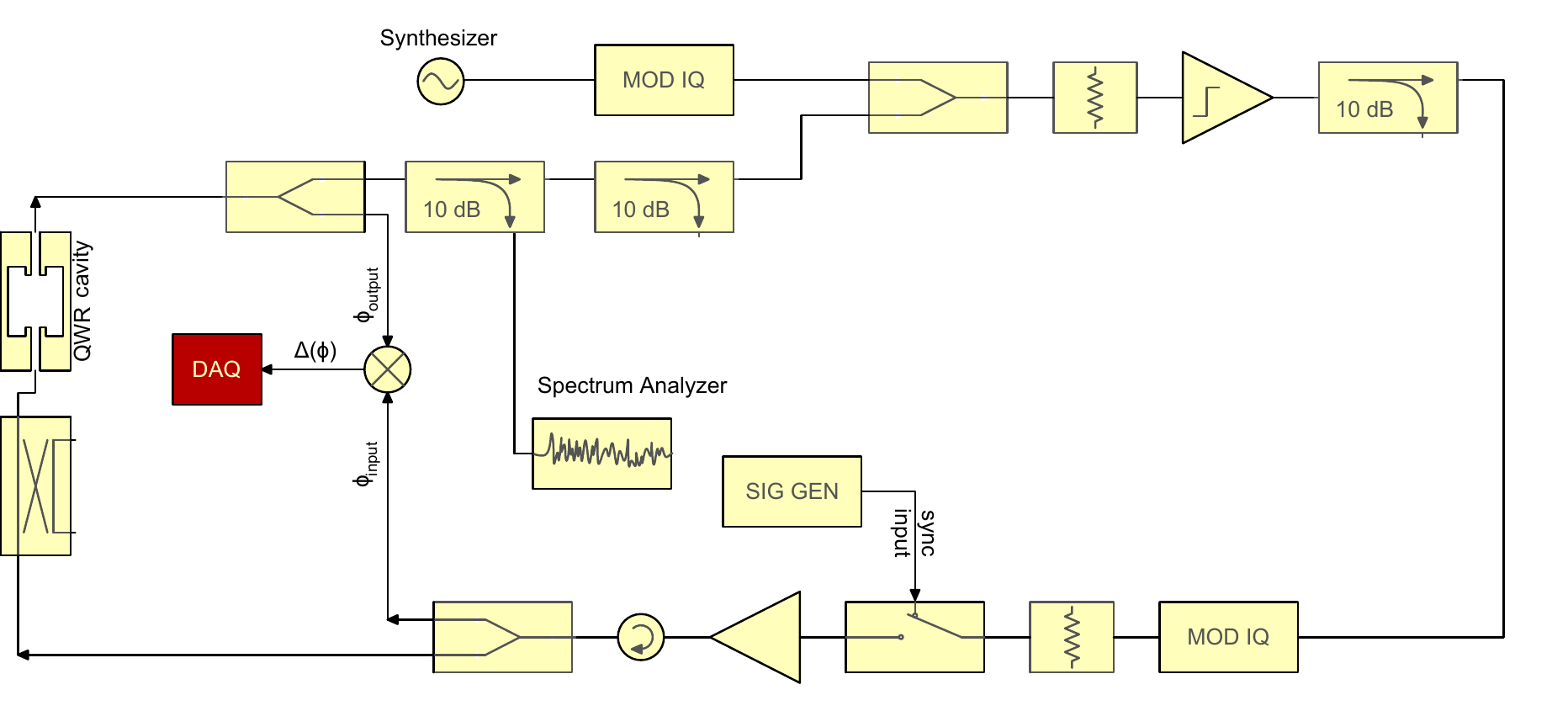}
  \caption{Block diagram of the RF measurement setup for the low power characterization the SPIRAL2 LINAC cavities.}\label{schema_bloc}
\end{figure}

In July 2017, we began the first injection of RF low power in the cavities for their cold characterisation \cite{GHRIBI2020103126}. We then noticed an amplitude modulation of the RF signal. Frequencies ranged between 4 to 6 Hz. The measurements were consistent and repeatable. However, resonance frequencies seemed to depend on the cavities positions. Figure \ref{spectre_rf} shows an example of such a spectrum. Figure \ref{schema_bloc} shows the diagram of the measurement test setup. A low power RF frequency generator was used to inject a signal locked at the resonance frequency of the measured cavity (found by a self-oscillating closed loop circuit). All measurements were done directly in the LINAC tunnel, bypassing the LLRF system to avoid the high losses that would have blinded us from seeing the signal we seek. The phase and the amplitude difference between the input and the output signal were extracted and measured thanks to a Yokogawa network analyser.

\begin{figure}[hbt]
  \includegraphics[width=.5\textwidth, right,trim={0cm 0cm 0cm 2cm},clip]{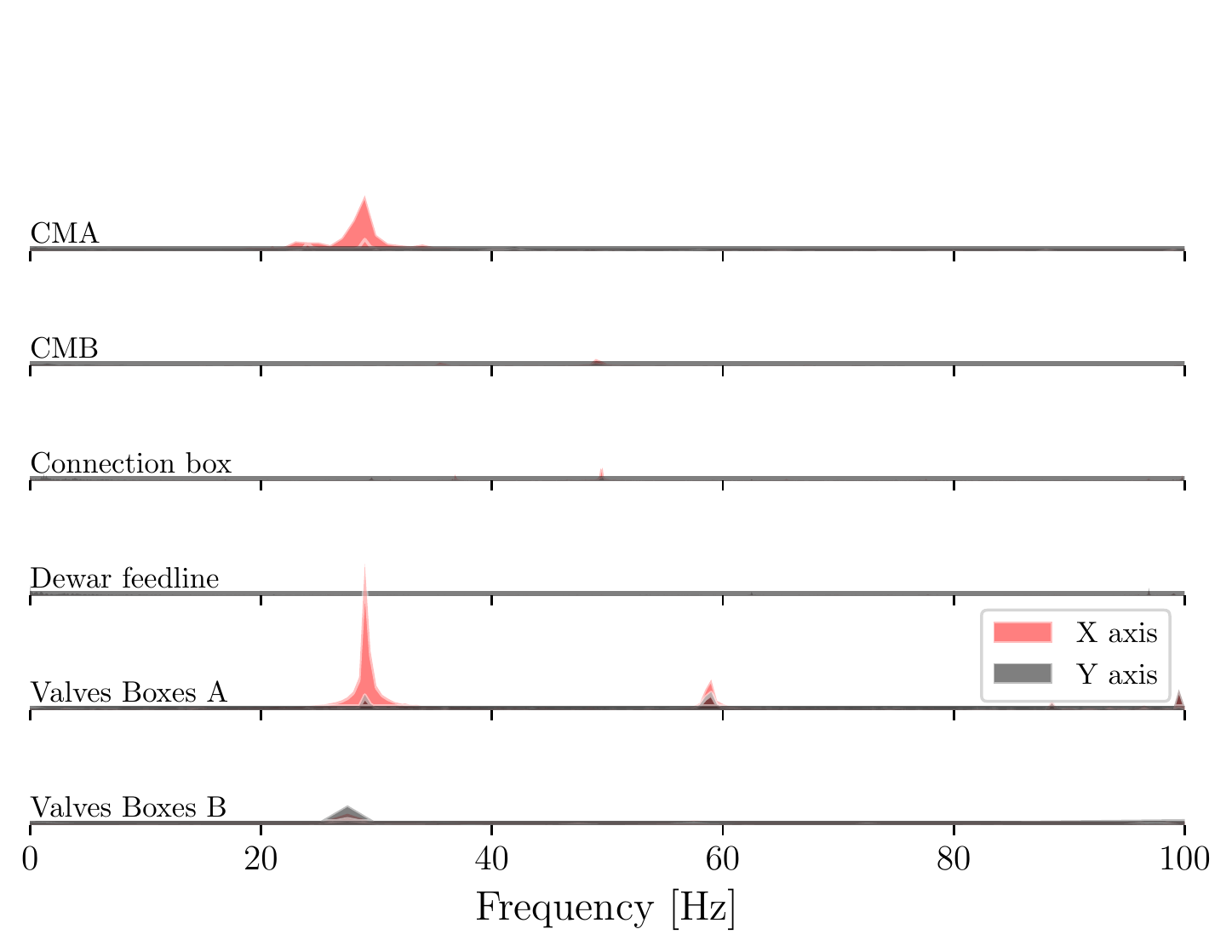}
  \caption{Accelerometer vibration measurements at different locations of the LINAC. A and B valves boxes are the valves boxes of cryomodules types A (CMA) and B (CMB). The X direction is the direction of the pipes. The Y direction is a horizontal orthogonal direction. The Z direction is not shown here but has not exhibit any peak. The peaks are normalized to an amplitude of G~=~0.1 are shown with the same scale.}\label{accelerometre}
\end{figure}

The first procedure that has been applied was external vibration measurements with a triple axis piezoelectric accelerometer. Different locations have been measured and monitored including : dewar pipes, valves boxes pipes, pumps and cryomodules. Figure \ref{accelerometre} shows a summary of these measurements. It has been shown that there is no correlation between the frequency of the roughing pumps and the measured peaks. It has also been found no peak bellow 10 Hz. However, peaks have been measured at 30 Hz and several of its harmonics only in the fluid direction (x direction in Fig. \ref{accelerometre}).

\section{Further experimental investigations}

\subsection{\label{sec_single_cm}Single cryomodule experiment}

\begin{figure}[hbt]
  \includegraphics[width=.4\textwidth, right,trim={3cm 3cm 4cm 4cm},clip]{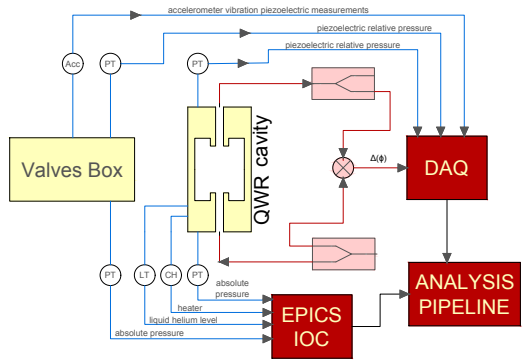}
  \caption{Block diagram of the 2018 measurement setup.}\label{schema_bloc2}
\end{figure}

\begin{figure}[hbt]
  \includegraphics[width=.5\textwidth, right,trim={0cm 0cm 0cm 0cm},clip]{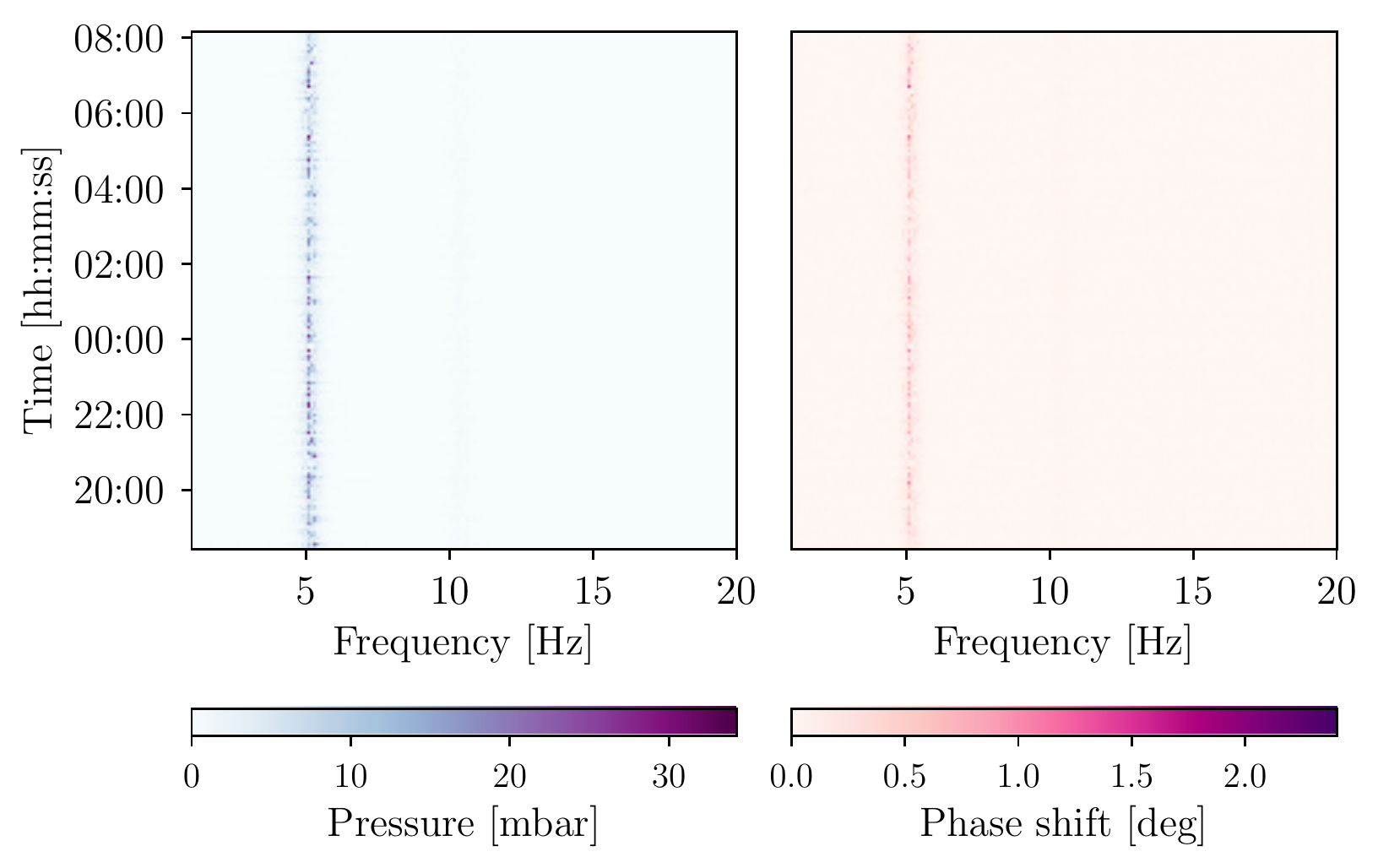}
  \caption{Overnight simultaneous measurements of pressure oscillations and RF phase shift modulations for cryomodule CMA04 (2018-11-24).}\label{phase_shift}
\end{figure}

In 2018, a different measurement setup has been installed. It included simultaneous measurements of absolute pressures, relative pressures, RF phase shifts, liquid helium level, heaters and accelerometers at different locations of a target cryomodule. A block diagram of such setup is shown in Figure \ref{schema_bloc2}. We measured both the absolute and the fast relative pressure directly in the cryomodule phase separator and in the corresponding valves box return line. We used piezoelectric sensors (PCB 113B28) for relative measurements and metallic process isolating diaphragm sensors (Cerabar PMP71) for absolute pressure measurements. We also measured vibrations with a tri-axis piezoelectric accelerometer. The same setup shown in Figure \ref{schema_bloc} has been used to extract the phase shift between the RF input and output signals. A National Instrument Compact DAQ centralised the fast acquisition with a 3 channels 2.8 kS/s/ch 24-Bit analog modules for the IEPE signals (Integrated Electronics Piezo-Electric) and a universal analog module for the RF signal. The NI DAQ was driven by an external laptop running Labview. Other data such as absolute pressures, heater power, liquid helium level and temperatures were measured through our regular PLC and archived with an Experimental Physics and Industrial Control System (EPICS) Input/Output Controller (IOC). A Python analysis pipeline assembled fast and slow acquisitions together with other correlation factors and a common clock. The results showed (see Figure \ref{phase_shift}) a direct correlation between the pressure measurements and the RF measurements. As a reminder, the purpose of this experiment is not only to understand but to attenuate this phenomena down to an acceptable level. The RF amplitude and phase variation were used as a proxy to determine the acceptable damping amplitude of the pressure oscillations. Given the margin on the RF amplifiers, the limit has been set to 20 dB in RF amplitude modulation which corresponds to roughly 10 mbars fast pressure variation.

\begin{figure}[hbt]
  \includegraphics[width=.5\textwidth, right,trim={7.5cm 1.5cm 6cm 2cm},clip]{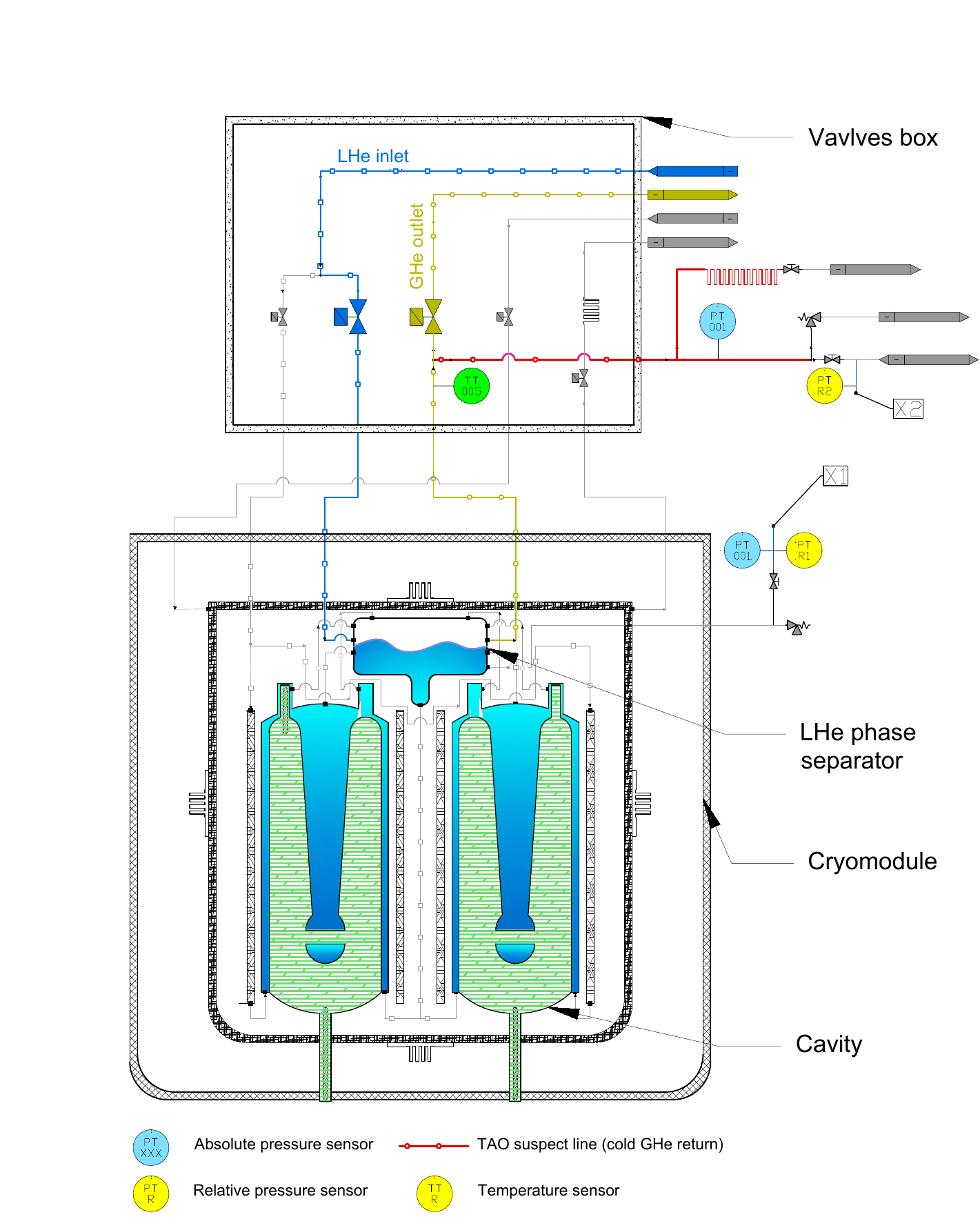}
  \caption{P\&ID of a type B cryomodule and its valves box. The red line is the low pressure return line from the LHe bath to the room temperature passive heater. Magenta circles are the main process valves. Yellow circles are the rapid relative piezoelectric pressure sensors. Blue circles are the absolute process pressure sensors. The green circle is the position of the temperature sensor on the return line.}\label{setup_tao}
\end{figure}

To better understand these measurements, a piping and instrumentation diagram of a type B cryomodule with its valves box is shown in Figure \ref{setup_tao}. The magenta coloured circles show the position of the two main process valves. The latters control the liquid helium level and the pressure in the cavity phase separator. Knowing that thermoacoustic oscillations usually appear for certain conditions with high temperature gradients, we investigated all room temperature ports with a cold end. We identified the red line (see Figure \ref{setup_tao}) that links the liquid helium bath to the security and purge ports of the valves box as a possible candidate. In order to investigate the possible start conditions of the oscillations (pressure, valves positions, thermal load, liquid helium level), we installed a couple of pressure sensors (absolute and relative) directly at the outlet of the security port of the liquid helium phase separator of the cryomodule (port \emph{X1}). We installed another one at the outlet of the candidate room temperature port (port \emph{X 2}). We found that the pressure oscillations start for a combination of pressure, liquid helium level and thermal load when the absolute pressure difference between \emph{X1} and \emph{X2} is around 1 mbar. We also noticed that when the TAO is present, the temperature of the return line increases. Thermodynamic modelling of the cryomodule and its valves box using the Simcryogenics library \cite{Bonne_2020} confirmed the consistency of the measurements with an increase of the thermal load in the return line that fitted the temperature difference of the return line before and after a TAO \cite{Vassal_2019}.

\begin{figure}[hbt]
  \includegraphics[width=.5\textwidth, right,trim={1.9cm 0cm 1.9cm 1cm},clip]{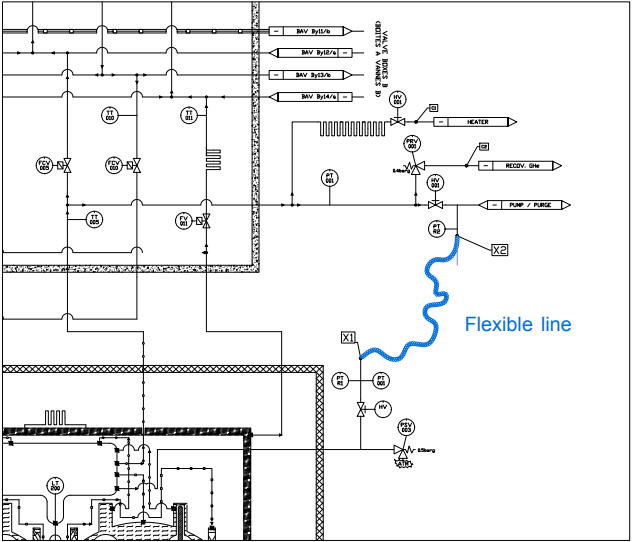}
  \caption{Vapor sky short circuit line solution setup - zoom in Fig. \ref{setup_tao}.}\label{setup_tao1}
\end{figure}

\begin{figure}[hbt]
  \includegraphics[width=.5\textwidth, right,trim={1.9cm 0cm 1.9cm 1cm},clip]{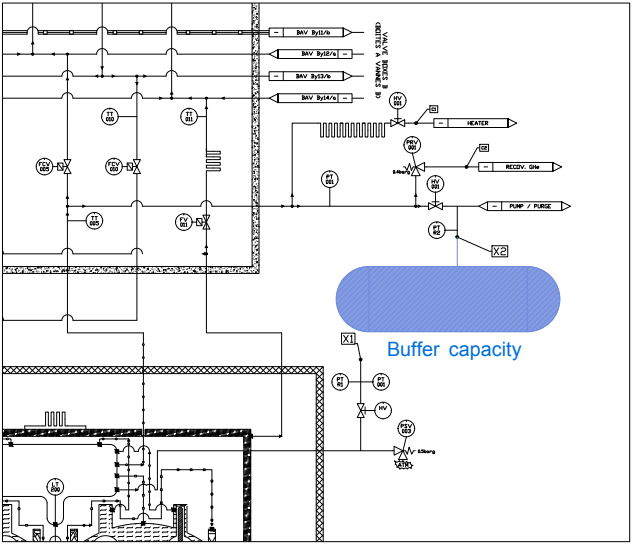}
  \caption{Buffer volume solution setup - zoom in Fig. \ref{setup_tao}.}\label{setup_tao2}
\end{figure}

\begin{figure}[hbt]
  \includegraphics[width=.5\textwidth, right,trim={1.9cm 0cm 1.9cm 1cm},clip]{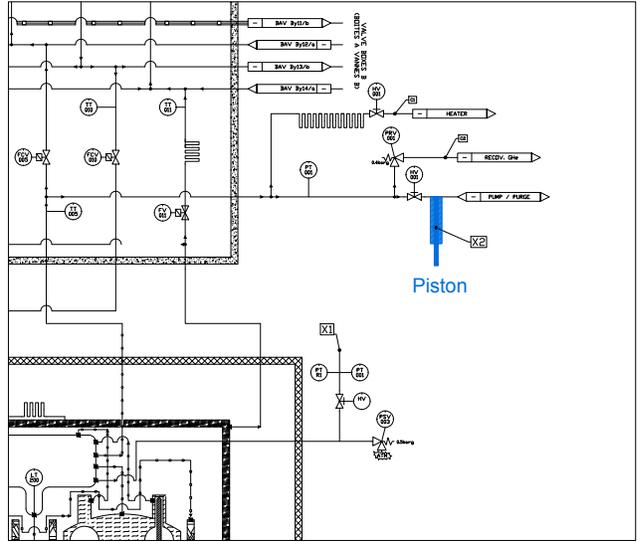}
  \caption{Piston solution setup - zoom in Fig. \ref{setup_tao}.}\label{setup_tao3}
\end{figure}

We then tried several methods documented in the literature \cite{GU1992194, ditmars1965, CHEN1999843, LUCK1992703} to suppress or damp the TAO :
\begin{enumerate}
	\item Short circuit between the phase separator vapor sky and the return line (see Fig. \ref{setup_tao1}) : Here we linked the two ports X1 and X2 with several pipes of several length and cross-sections.
	\item Buffer (see Fig \ref{setup_tao2}) : Here we connected several buffers of different volumes to the port X2.
	\item Piston (see Fig. \ref{setup_tao3}) : Here we inserted a piston in the port X2 and we monitored the behaviour of the system for several insertion depths.
\end{enumerate}

For every solution tried, we spanned all operating conditions to determine the most suitable solution for our case. In order to compare the results, we used an efficiency criteria defined as :
\begin{equation}
	\zeta = \frac{P_{bath}^{off}}{P_{bath}^{on}}
\end{equation}
where $P_{bath}^{off}$ is the amplitude of the pressure oscillations with no TAO correction and $P_{bath}^{on}$ is the amplitude of the pressure oscillations with the considered TAO correction.

\begin{figure}[hbt]
  \includegraphics[width=.5\textwidth, right,trim={0cm 0cm 0cm 0cm},clip]{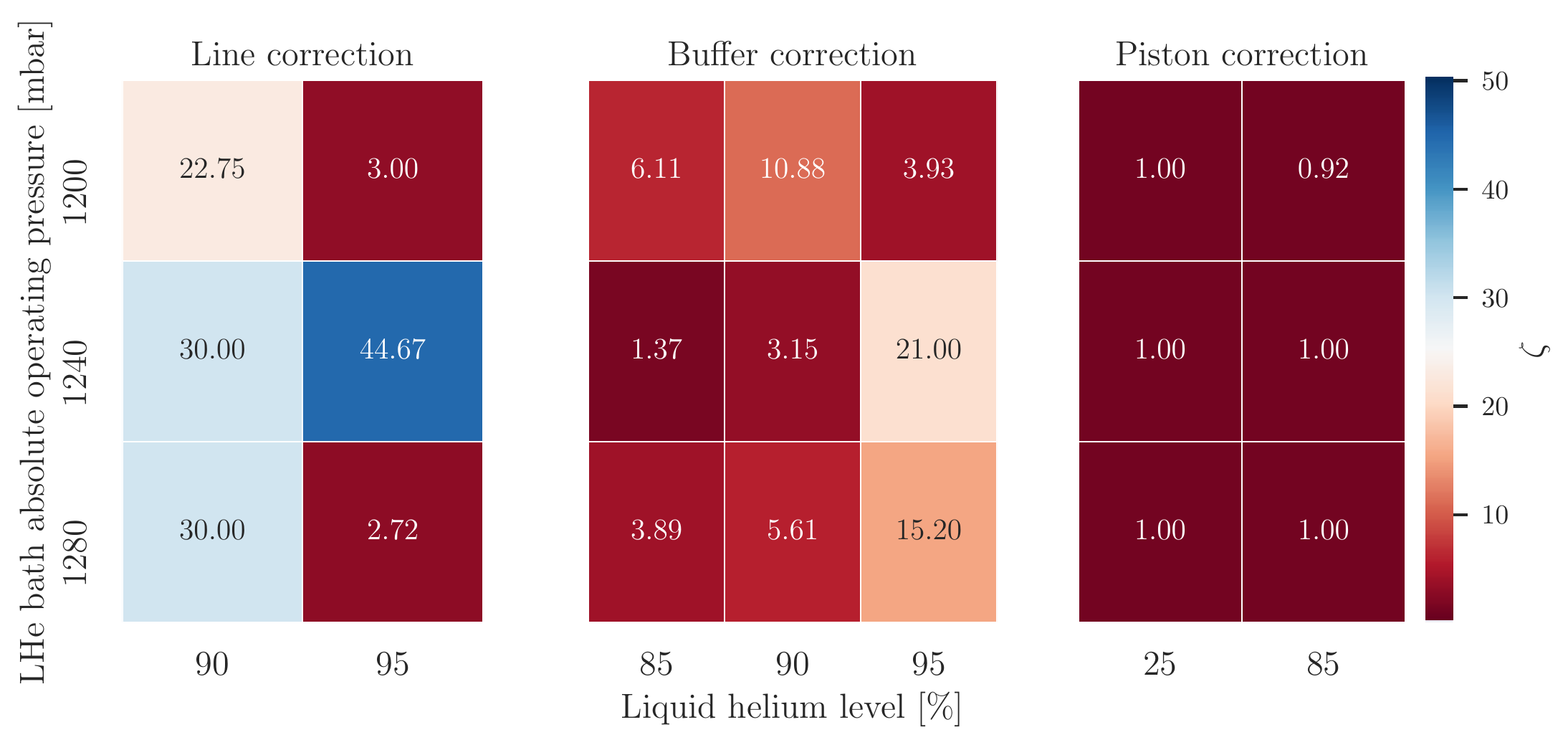}
  \caption{TAO correction efficiency $\zeta$ as a function of the pressure (PT001) and the liquid helium level (LT200) for the three considered experimental corrections : line (vapor sky short circuit), buffer and piston for cryomodules CMA05 and CMB03.}\label{eff_taoc}
\end{figure}

For every applied correction, we did see an effect on TAO damping but no total suppression has been achieved. An example of efficiency of every correction is shown in Figure \ref{eff_taoc} for different pressures and liquid helium operating conditions. It appeared that the most efficient solution for every case is the line short circuit correction. This solution was efficient enough to be deployed in all LINAC. The TAO correction efficiency reached 1000 for some cases, damping the oscillations from [10-100] mbars to les than 0.1 mbar. The short circuit line was sufficient to recover the pressure balance between port X1 and port X2, therefore avoiding the appearance of high amplitude TAO oscillations. However, the flow rate was so important sometimes that it froze part of the line and the upper neck of the cryomodule or resulted in condensation in the same locations. We then deployed it to all cryomodules with permanent lines terminated by a on/off hand valve for one end, a micro-metric hand control valve for the other hand and pressure security valve in between. The micro-metric valve allowed us to control the flow in the correction line in order to avoid water condensation or ice. The on/off valve allowed us to suppress any flow through the correction line. This was useful especially when cooling down the LINAC. This on/off valve allowed us also to de-activate or activate the TAO correction for a more thorough investigation of TAO amplitudes and frequencies cross-couplings.

\subsection{Multiple cryomodules investigation}

While it might seem obvious, for an isolated system, that thermoacoustic oscillations occur because of local conditions, the answer is not clear for coupled or connected systems. For the LINAC, we are in the complex case of interconnected cryogenic clients with several room temperature ports with cold ends. It is therefore unclear wether the amplitude and frequency dependance of the oscillation is dominated by local effects or global effects. Observed transient pressure fluctuations with sudden changes in frequency of the TAO could be due to such connections. 

\subsubsection{Experiment description}

To have a better picture, we repeated the experiment described in subsection \ref{sec_single_cm}  and deployed the setup described in Figure \ref{schema_bloc2} in all the LINAC. We deployed 19 piezoelectric relative pressure sensors (one for every cryomodule) in the LINAC and one at the central cold box return line. Acquisition was made by the same DAQ  previously described with seven 24-Bit analog modules controlled with a Labview program. As previously, all fast acquisition data were treated with a python pipeline analysis that combined PLC slow sensors and NI DAQ fast sensors. This time, the pipeline allowed automatic peak extraction and TAO detection. This allowed to have a much better view on what was going on in the LINAC. Several configurations have been tested :

\begin{description}
	\item Configuration 1 \textbf{Full LINAC OFF} \\ In this configuration, no TAO correction has been applied. This configuration was meant to see the distribution of amplitudes and frequencies of the TAO in the LINAC with mixing individual cryomodules contributions as well as crosstalks.
	\item Configuration 2 \textbf{Full LINAC OFF-1} \\ In this configuration, TAO correction has been applied to only one cryomodule at a time. That means that we closed all short circuit lines valves in all the LINAC, disabling the TAO correction but in one cryomodule. We then spanned this configuration over all cryomodules. This setup was meant to see the effect of other cryomodules that had an activated TAO on a cryomodule with an activated correction.
	\item Configuration 3 \textbf{Full LINAC ON} \\ In this configuration, we activated the TAO correction in all cryomodules at the same time. This was meant to observe if we had other fast transient pressure fluctuations.
	\item Configuration 4 \textbf{Full LINAC ON-1}\\  In this configuration, TAO correction was activated in all LINAC but in one cryomodule at a time. We then spanned this configuration to all cryomodules. This was meant to observe the frequency and amplitude of a local TAO correction and if it had an effect on other cryomodules.
\end{description}

\begin{figure}[hbt]
  \includegraphics[width=.5\textwidth, right,trim={0cm 0cm 0cm 0cm},clip]{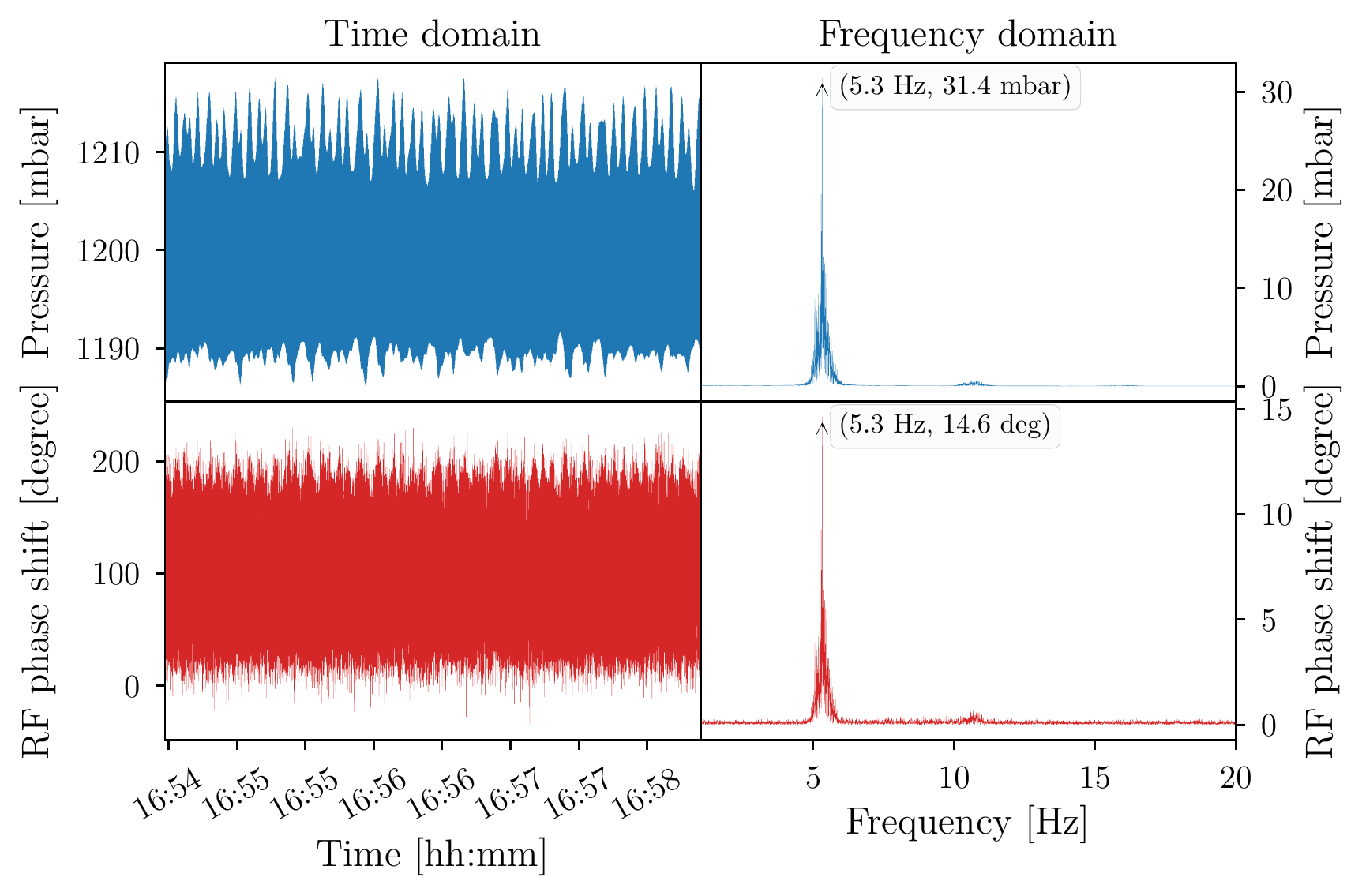}
  \caption{Example of time and frequency domain data extracted and calculated with the pipeline analysis for CMA11 [2018-11-11].}\label{timef}
\end{figure}

\subsubsection{Results\label{results}}
\begin{figure}[hbt]
  \includegraphics[width=.5\textwidth, right,trim={0cm 0cm 0cm 0cm},clip]{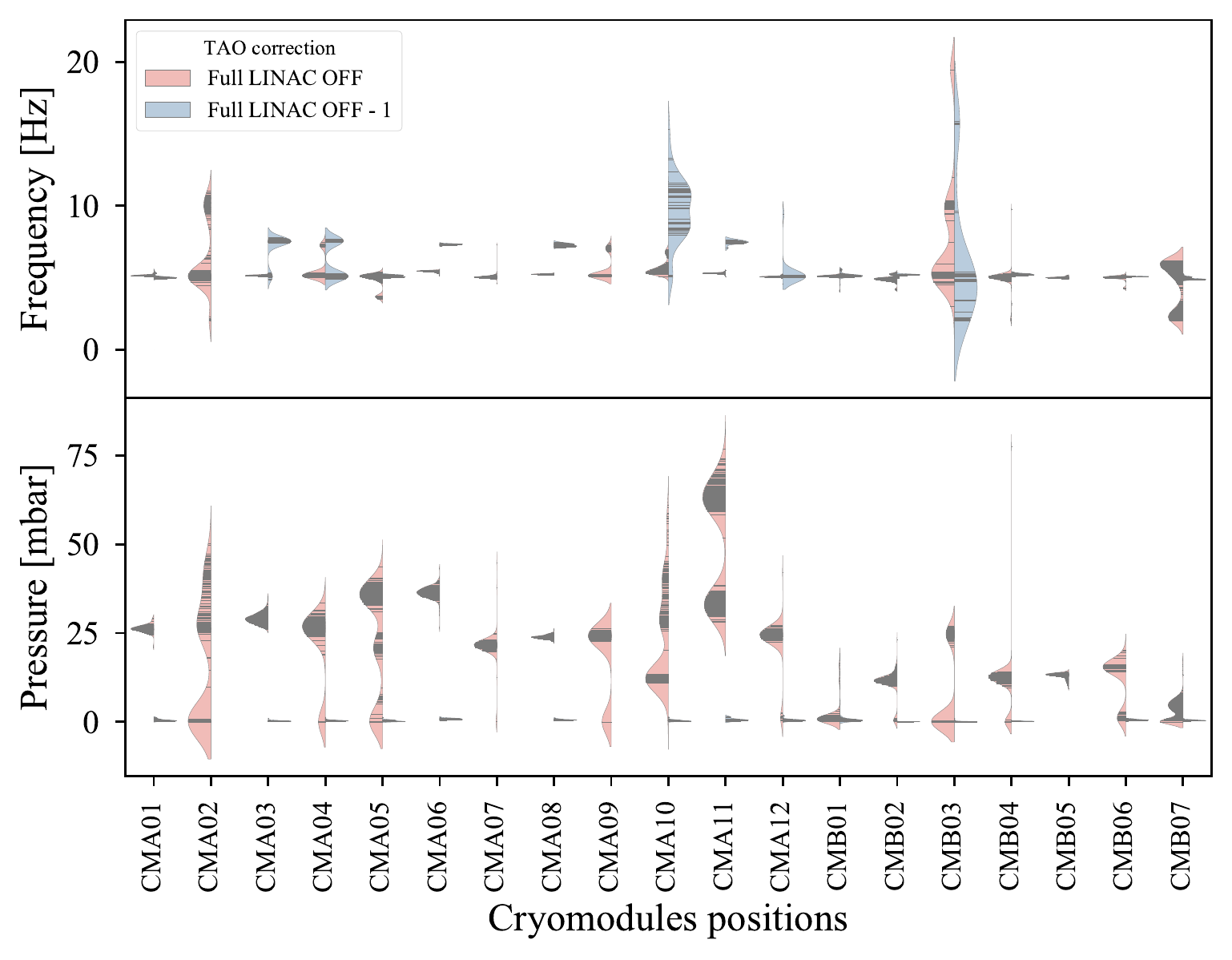}
  \caption{Full LINAC OFF vs Full LINAC OFF-1 : kernel distributions half violin plots normalized at widths for comparison between configuration 1 and configuration 2.}\label{offoff-1}
\end{figure}

\begin{figure}[hbt]
  \includegraphics[width=.5\textwidth, right,trim={0cm 0cm 0cm 0cm},clip]{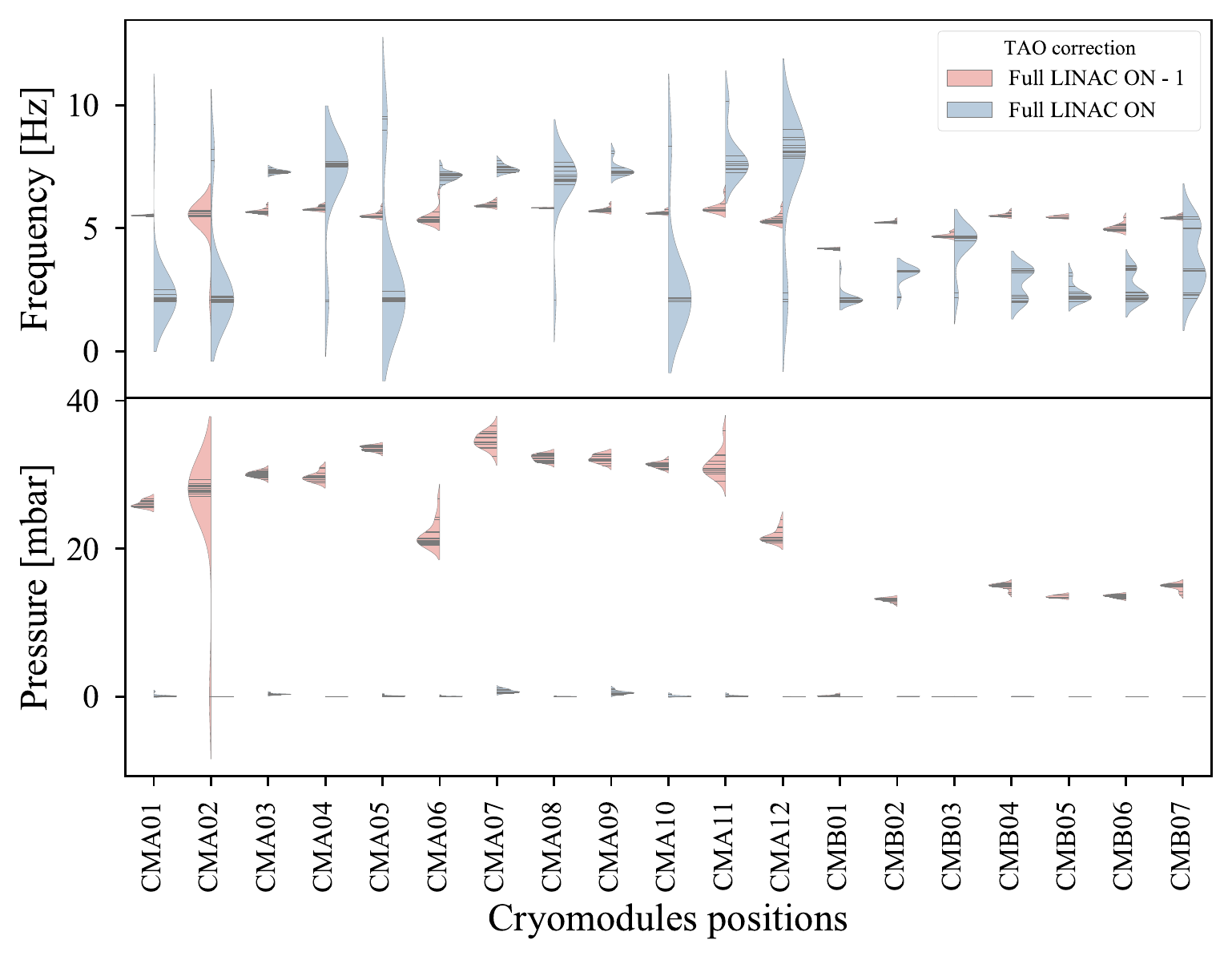}
  \caption{Full LINAC ON vs Full LINAC ON-1 : kernel distributions half violin plots normalized at widths for comparison between configuration 3 and configuration 4.}\label{onon-1}
\end{figure}

\begin{figure}[hbt]
  \includegraphics[width=.5\textwidth, right,trim={0cm 0cm 0cm 0cm},clip]{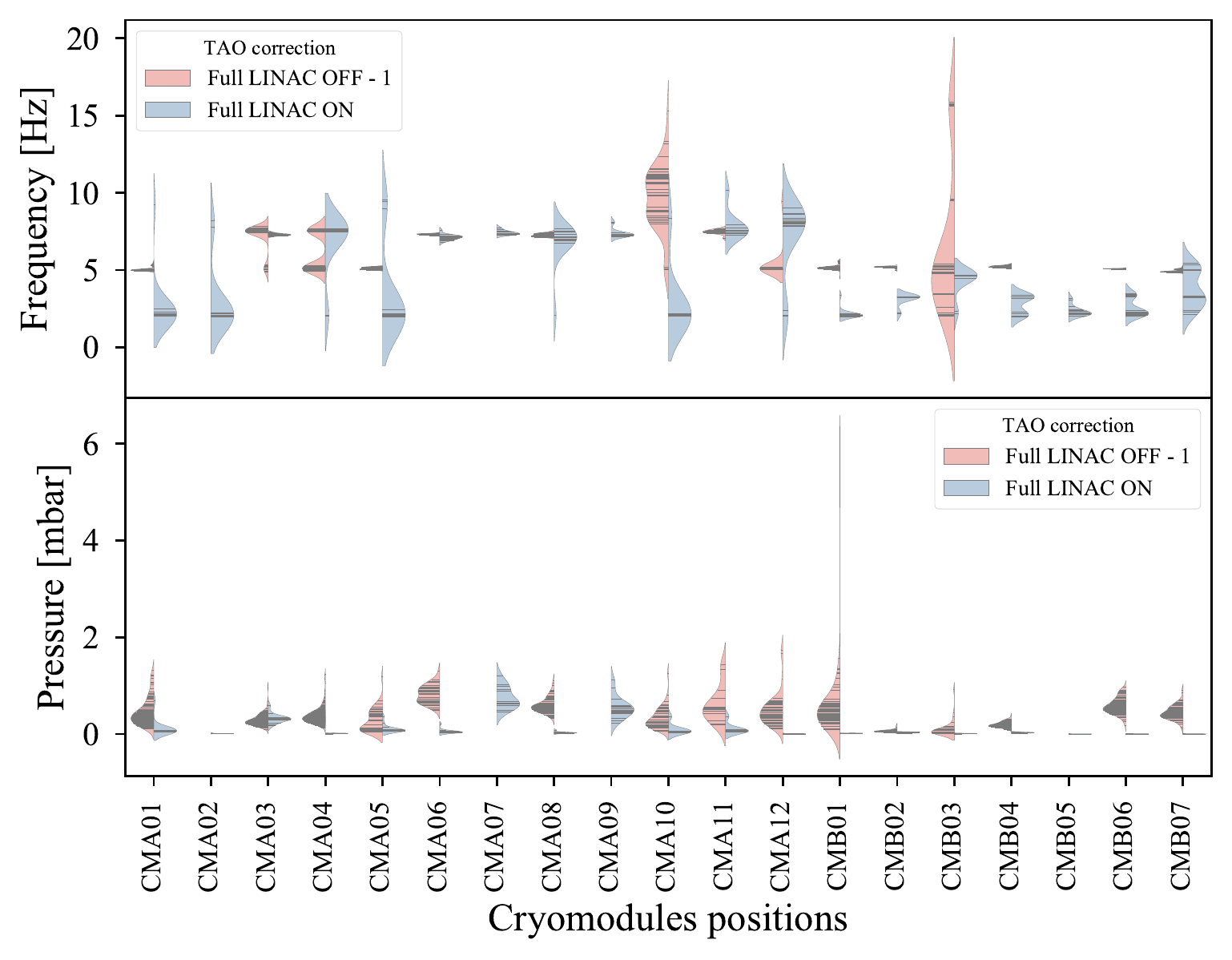}
  \caption{Full LINAC OFF-1 vs Full LINAC ON : kernel distributions half violin plots normalized at widths for comparison between configuration 2 and configuration 3.}\label{off-1on}
\end{figure}

As described previously the pipeline allowed automatic extraction of both frequency and amplitudes of the TAO oscillations. All data were gathered in time domain. The reference of a relative pressure sensor is always zero which means that the correct amplitude is found by offsetting the relative data by the absolute pressures measured with the CERABAR pressure sensors. The amplitude of the oscillations was found by enveloppe calculations within the considered time window using its Hilbert transform. For the frequency peak detection, we first applied a Fourrier transform on the data in a time window of 4 minutes in order to have a high resolution. We then applied a high pass filter to avoid the 1/f noise bellow 1 Hz and a low pass filter to avoid high order harmonics. We finally computed the centroid of the resulting spectrogram to extract the frequency peak. An example of a time and frequency domain extracted data is shown in Figure \ref{timef}. Thanks to the extracted data, Parzen-Rozenblatt kernel density estimations have been computed for both frequencies and amplitudes of TAO oscillations for every considered configuration. The analysis of 10,787 datasets allowed the comparison of the different configurations :

\begin{description}
	\item \textbf{Full LINAC OFF} VS \textbf{FULL LINAC OFF-1} Fig. \ref{offoff-1}\\
		\begin{itemize}
		\item[\emph{Frequency}] For the configuration 1, with no TAO correction, we see that most cryomodules have a very sharp resonance frequency around 5 Hz. There are several exceptions such as CMA02, CMA04, CMA10 and CMB03 that exhibit multiple peaks. When the single cryomodule TAO correction is applied  (configuration 2) we see a frequency shift for most cryomodules.
		\item[\emph{Amplitude}] For the configuration 1, we see that the TAO amplitudes are distributed unevenly. The wider distributions concerns most type A cryomodules. We also see that the single cryomodule TAO correction is very efficient as all amplitudes drop to almost 0 mbar. No effect of the remaining TAO oscillations is visible here.
		\end{itemize}
	\item \textbf{Full LINAC ON} VS \textbf{FULL LINAC ON-1} Fig. \ref{onon-1}\\
	\begin{itemize}
		\item[\emph{Frequency}] This time, when only one cryomodule at a time has an activated TAO, the oscillation frequencies seem un-correlated. When the correction is applied to all the LINAC, we see many low amplitude oscillation frequencies with a distribution that seem common to the type A cryomodules and a distribution that seem common to the type B cryomodules. This suggests the native presence of multiple low amplitude and geometry dependent oscillations occurring in the LINAC.
		\item[\emph{Amplitude}] When only one TAO is activated at a time, pressure amplitudes seem less widely distributed than in the configuration 1. This suggests mixed contributions of distributed TAO oscillations in the configuration 1. Once again, amplitudes seem to be geometry dependent as more similarities arise among type A cryomodules and among type B cryomodules. The efficiency of the TAO correction (configuration 3) is noticeable as we see no high amplitude TAO occurence when the correction is activated for all LINAC.
	\end{itemize}
	\item \textbf{Full LINAC OFF-1} VS \textbf{FULL LINAC ON} Fig. \ref{off-1on}\\ This comparison asks specifically the question of the effect of neighbouring or distant TAO oscillations on a cryomodule that has an activated TAO correction.
	\begin{itemize}
		\item[\emph{Frequency}] When other TAO oscillations are present, most of the time, we see clear sharp single or double peak frequencies. These are signatures of distributed resonances and do not include a local resonance (damped by the short circuit line). When the remaining TAO are damped, frequencies distributions take a wider shape. The latter conclusion is certainly a generalisation that does not include CMA10 et CMB03 that exhibit double resonance like behaviour. Nor does it include CMA06 and CMA07 and CMA09 that keep a sharp single peak resonance.
		\item[\emph{Amplitude}] The observations of the frequency distributions are confirmed by the amplitudes distributions. The latters drastically drop when all TAO corrections are activated. CMA07 and CMA09 are exceptions that remain and show both sharp frequency peaks and noticeable resonance amplitudes (can reach 1 mbar). This is not critical as the phase modulation or rather RF impedance modulation resulting in these oscillations is not significant and can be compensated by the RF amplifiers power margin.
	\end{itemize}
\end{description}

\section{Analytical investigation}
\begin{figure}[hbt]
  \includegraphics[width=.5\textwidth, right,trim={0cm 0cm 0cm 0cm},clip]{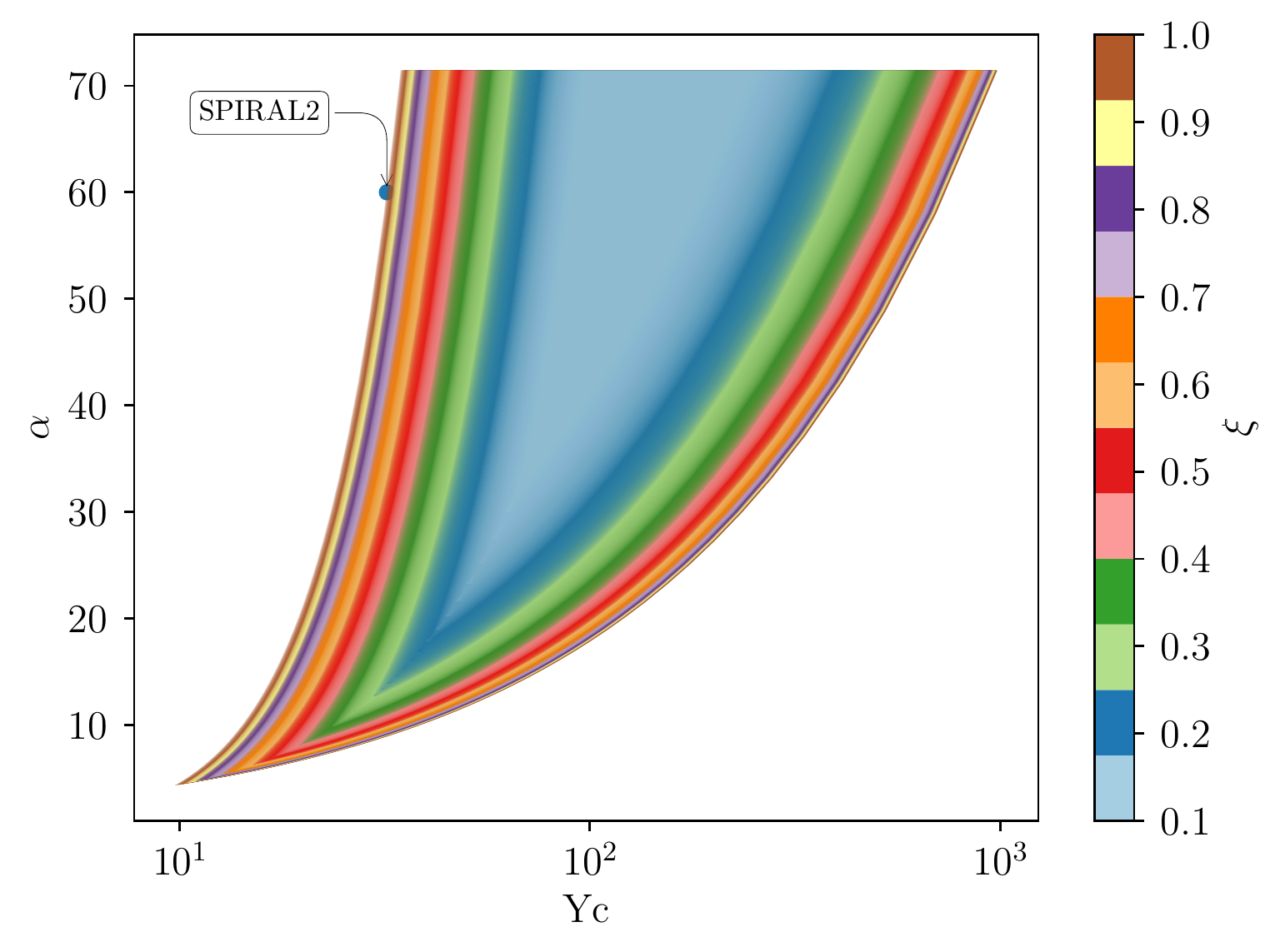}
  \caption{Rott stability curve and the case of SPIRAL2.}\label{rott}
\end{figure}

Given the known inner dimensions of the cryogenic pipes, the volumes and the temperatures, we tried an analytical investigation of a localised thermoacoustic oscillation in a cryomodule. We used Rott's theory \cite{ROTT1980135} to investigate criteria that drives the appearance of thermoacoustic oscillations. We now know that being able to describe these phenomena heavily depends on our understanding of the viscous actions of the high density helium in the inner tubes surfaces. We also know that these depend on several geometric parameters such as tubes diameters, lengths, thicknesses as well as materials and temperature gradients. To better represent the boundary conditions of appearance of the TAO, Rott has introduced a dimensionless quantity $Y_{c}$ that represents the ratio of the inner tube radius to the Stokes boundary layer thickness. $Y_{c}$ also drives the amplitudes of the Taconis oscillations. The Rott's approach is a stepped stability analysis with a discontinuity model where a temperature jump happens somewhere in the tube where the oscillation occurs \cite{rott1973}. The position of the temperature jump is described as follows :
\begin{equation}
	\xi = \frac{L-l_{c}}{l_{c}}
\end{equation}
with $L$ the length of the tube between the cold end and the warm end and $l_{c}$ the length at which the temperature jump occurs (from the cold end). $Y_{c}$ is then given by :
\begin{equation}
	Y_{c} = \frac{2D\alpha^{1+\beta}}{a+\xi^{-1}+\lambda_{c}^{2}\xi}
\end{equation}
with $\alpha$ being the ratio between the temperature of the warm end $T_{H}$ and the cold end $T_{C}$ :
\begin{equation}
	\alpha = \frac{T_{H}}{T_{C}}
\end{equation}
Here, additional constraints apply on $\alpha$ depending on $\xi$. When $0<\xi<1$, $\alpha = 0.228~\xi~Y_{c}^{2.214}$. When $\xi > 1$, $alpha=0.0264~Y_{c}^{2.429}$.
$\lambda_{c}$ is a dimensionless frequency parameter defined by :
\begin{equation}
	\lambda = \frac{\omega~l}{a_{c}}
\end{equation}
where $\omega$ is the angular frequency, $a_{c}$ is the speed of sound in the gas at the cold end, $l$ is the tube cold length.
Usually, an assumption is taken that $\xi = 1$ meaning that the temperature jump occurs in the middle. In our case we will consider all the solutions with $\xi$ varying from 0 to 1. 
To search for a solution for $\lambda_{c}$, we can consider the equation from \cite{GUPTA2015104} :
\begin{equation}
	Y_{c}\lambda^{\frac{1}{2}} = \left( \frac{a_{c}}{\lambda_{c}} \right)^{\frac{1}{2}}r_{0}l^{-\frac{1}{2}}
\end{equation}
where $\nu_{c}$ is the kinematic viscosity of the the fluid. This leads to :
\begin{equation}
	Y_{c}\lambda_{ c}^{\frac{1}{2}} = r_{0} \left( \frac{a_{c}L}{\nu_{c}(\xi + 1)}\right)^{\frac{1}{2}}
\end{equation}
For helium at such temperatures, we consider $a_{c} = 1.2~10^{4}~cm.s^{-1}$. Given Rott's analysis, D and $\alpha$ are constants with values that depend on the intermolecular force field \cite{GUPTA2015104}. For helium between $300~K$ and $4.5~K$ they are set to $D=1.19$ and $\beta = 0.647$. For our case, we consider also the inner radius of the tubes to be $r_{0} = 9~mm$. These equations lead us the two branches of the Rott's stability curve shown in Figure \ref{rott}. This plot spans all cases of $\xi$ from 0 to 1. The outer surface of the trace represents the stability region whereas the inner surface represents the region where a thermoacoustic oscillation can occur. Given the experimental observations, we also consider that $T_{C}=5~K$ and $T_{H}=300~K$, which leads to $\alpha = 60$. This makes the SPIRAL2 case fall at the edge of the stability curve implying that the overall length of the line is smaller than one meter. Given the observations (frequencies and geometries), it is safe to say that the estimations of the location of the SPIRAL2 case in the Rott's stability curve is not correct. Several assumptions can lead to this mistake. In fact, the geometry of the cryomodule/valves box ensemble is not as simple as a simple pipe. For example depending on the position, the inner tube has different diameters. The immediate environment also has an effet on the conditions of oscillations kick-off. It is therefore necessary to have more detailed approach that fits the complex situation of inclusive systems where cryogenic thermoacoustic oscillations occur.

For the moment, let us consider that we don't know anything about the effective geometric parameters that would allow us to make a Rott's analysis. Let us consider, as did Rayleigh in 1877 \cite{KU2005588}, the system as an oscillator (which in fact it is). In this situation, the gas in the critical volume V is the spring mass and the gas in the tube is the proving mass. The frequency of the oscillation, can be directly derived from eq. 3.48 in \cite{ROTT1980135} as :
\begin{equation}
	\omega = \left( \frac{\pi r^{2} l_{c}}{V} \right)^{\nicefrac{1}{2}}
\end{equation}

If we consider that the oscillating mass is the mass in the tube, with $V=\pi r^{2} (L-l_{c})$ and $\xi = (L-l_{c})/l_{c}$, then it comes :

\begin{equation}
	l_{c} = \frac{a_{c}}{2\pi \nu \xi^{-1}}
\end{equation}

\begin{figure}[hbt]
  \includegraphics[width=.5\textwidth, right,trim={0cm 0cm 0cm 0cm},clip]{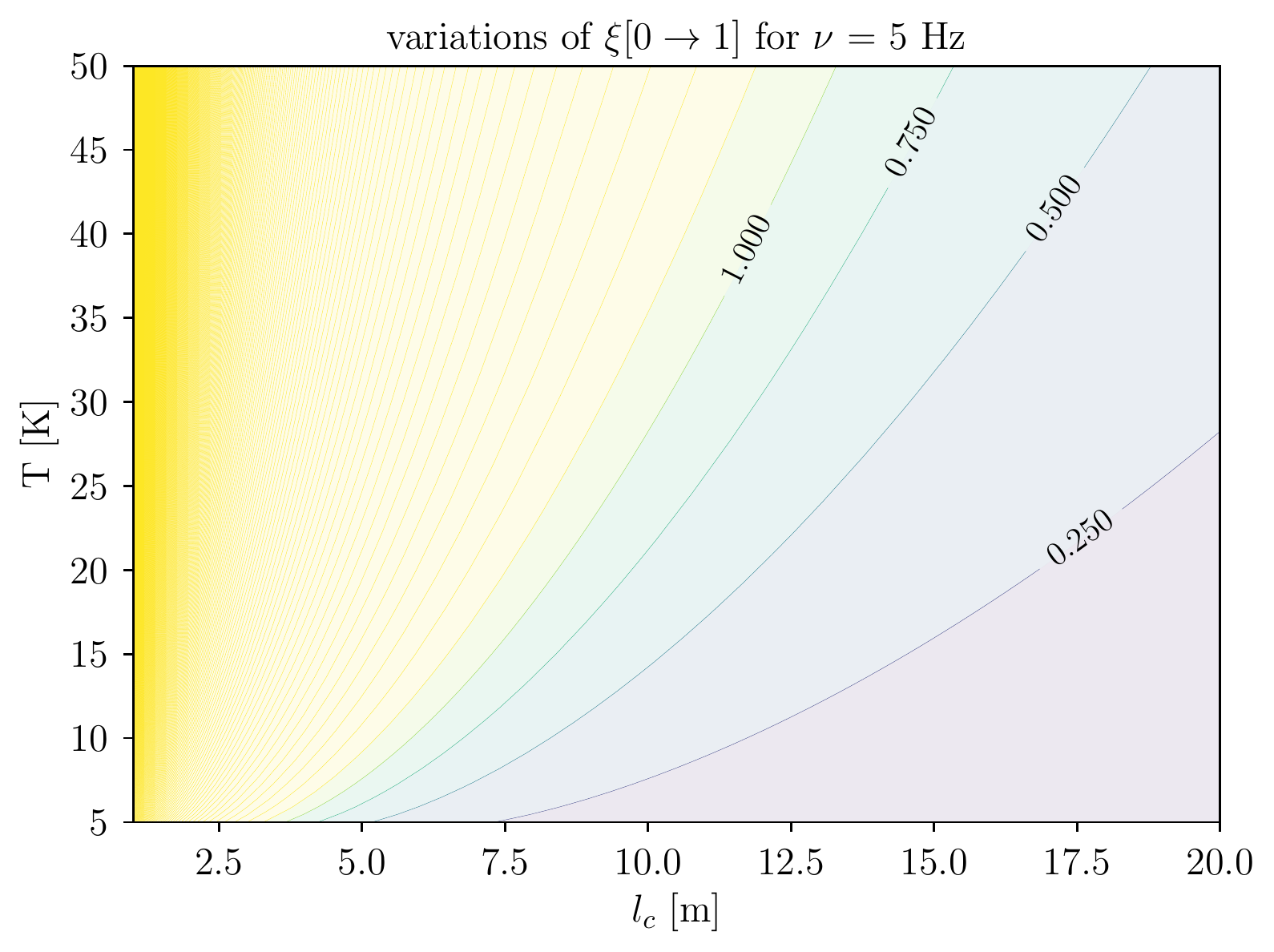}
  \caption{Contours of the cold effective cold end length as a function of the cold temperature value for different values of $\xi$ and $\nu = 5$ Hz.}\label{nu_lc}
\end{figure}

\begin{figure}[hbt]
  \includegraphics[width=.5\textwidth, right,trim={0cm 0cm 0cm 0cm},clip]{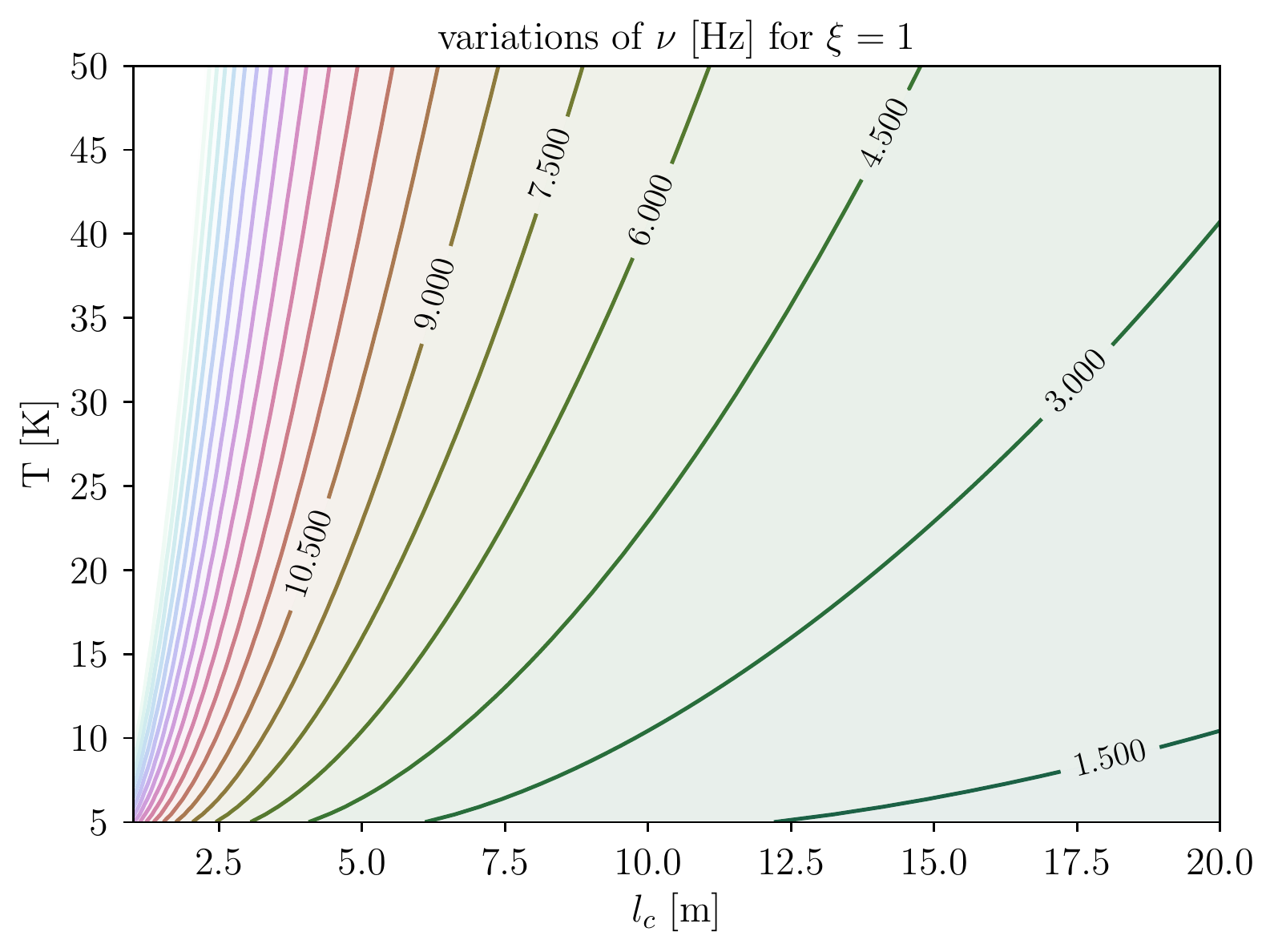}
  \caption{Contours of the cold effective cold end length as a function of the cold temperature value for different values of $\nu$ and $\xi = 1$.}\label{nu_lc2}
\end{figure}

We have observed that when the single oscillators are coupled, we are dominated by a central frequency lying around 5 Hz (see Figure \ref{offoff-1}). If we consider this as an input and vary the effective cold end length and the temperature of the cold end as a function $\xi$, we see (Figure \ref{nu_lc}) that, in order to have small cold ends, we need $\xi \geq 1$. On the other hand we know that the oscillators are all different and exhibit different frequencies ranging between 4 and 10 Hz. If chose $\xi = 1$ (which is usually considered for this kind of analysis), we see from Figure \ref{nu_lc2} that the effective cold end length should be somewhere between $2 m$ and $15 m$. If we consider an electromagnetic analogy and given the low frequencies observed (with respect to acoustic frequencies), this fits much better with a long wavelength. The problem that arises again here is that the effective impedance of an interconnected system is not easy to estimate. This is critical in order to be able de design a proper damper or any other solution to this kind of problem. It is worth mentioning here that, in these considerations, the dynamic behaviour of He has been taken into account with respect to its temperature and the pressure of the process (1300 mbars). Calculations have shown that there is very little dependance between the frequency of the oscillations and the pressure of the process. However, experiments have showed a high dependance between the pressure and the start up conditions of the TAO.

\section{Damper design}

The correction that proved efficient in subsection \ref{results} has few drawbacks. It creates a helium flow in the pipe passage where the liquid helium level sensor stands. This modifies the cold helium stratification shape in the liquid helium pipe volume and hence affects its behaviours. It manifests as a plateau in the liquid helium readout. This effect has been observed several times during operation. The plateau's location was different for every cryomodule but it was usually between 89\% and 97\%. Such plateau is a non linear behaviour that can't be handled by the PID controller in charged of the level control. In the end the global stability of the liquid level is affected by those plateau. One explanation is that, in the upper part of the level probe, the additional helium flow causes a sub-cooling of the level probe superconducting wire, which maintains the wire in the superconducting state independent of the helium level. Tuning the current and the boost current of the liquid helium level sensors proved inefficient to solve the problem. We therefore spanned a range of set-points for liquid helium level control in order to find a suitable set-point compromise. Although this compromise has been found for our current operation conditions, it is suitable to find a different solution that does not affect the behaviour of a critical component of the cryostats. In subsection \ref{results}, we used what we had in hand to solve quickly and efficiently the problem. A proper design of a suitable damper would solve the problem.

In \cite{LUCK1992703}, instructions are given on how to design such a damper. Let's explore the buffer reservoir solution. If we consider the incoming wave impedance
\begin{equation}
	Z_{i} = \frac{\rho a}{A}
\end{equation}
, the reservoir impedance
\begin{equation}
	Z_{r} = \frac{-i \rho a^{2}}{\omega V}
\end{equation}
and the orifice impedance (valve) :
\begin{equation}
Z_{0} = \frac{\sqrt{8\mu\rho\omega}(1+h/2r+\Delta)+i\rho\omega(h+1.7r)}{r^2 \pi}
\end{equation}

the wave is damped when :
\begin{equation}
Z_{i}=Z_{r}+Z_{0}
\end{equation}

\begin{figure}[hbt]
  \includegraphics[width=.5\textwidth, right,trim={0cm 0cm 0cm 0cm},clip]{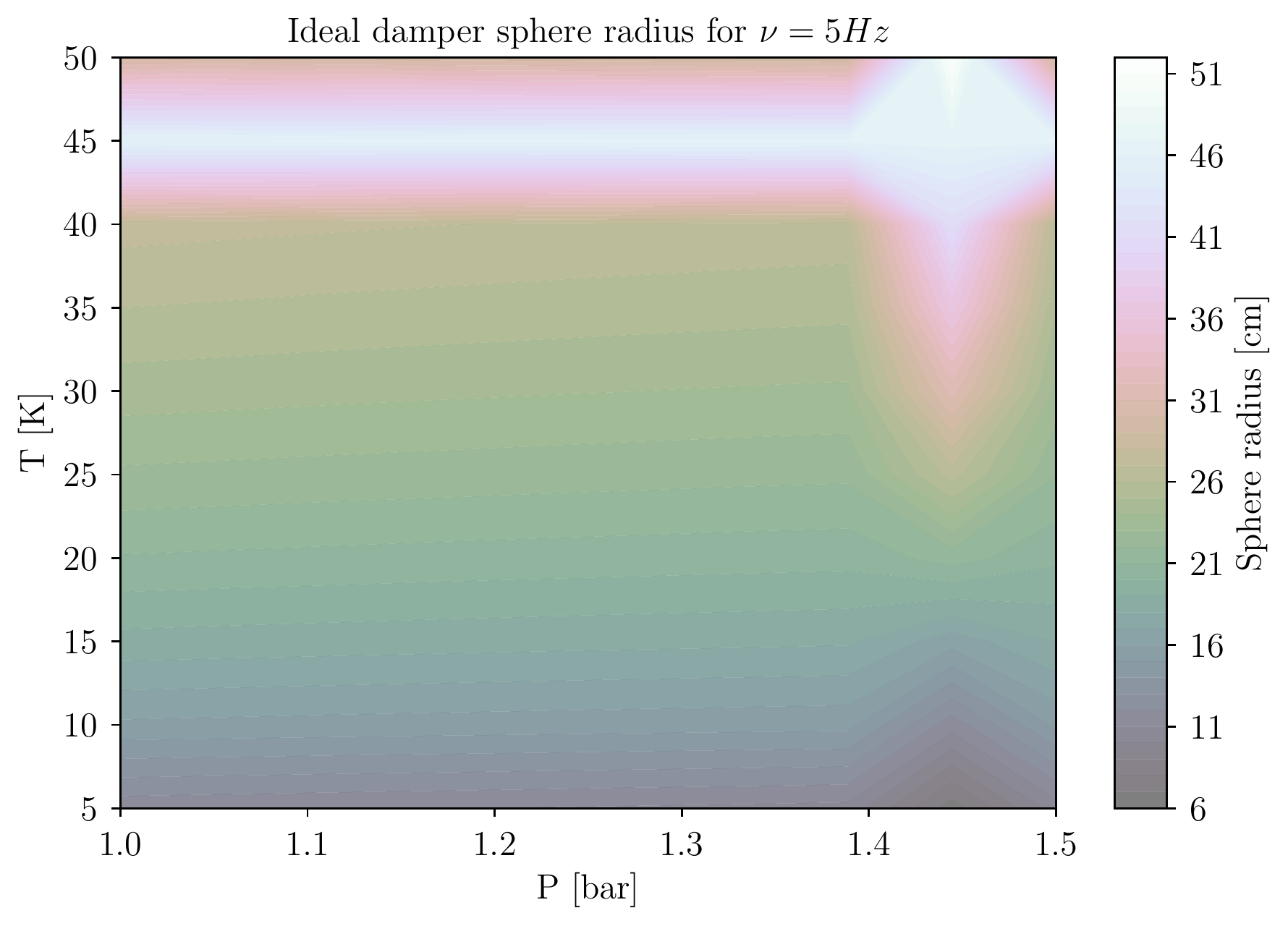}
  \caption{Damper ideal sphere diameter as a function of the temperature and the pressure of He gas for an oscillation at 5 Hz.}\label{damper}
\end{figure}

Obviously, the solution depends on the temperature of the gas and its pressure as well as the wave parameters. Figure \ref{damper} shows a design compromise contour for different gas temperatures and pressures for the geometrical parameters of the SPIRAL2 valves boxes and the dimensions of the port considered for the correction.

\section{Conclusion}
Investigating cryogenic thermoacoustic oscillation occurring in a tube in a lab experiment and a real life TAO in a working superconducting LINAC is a completely different thing. Although the two share the same physical principles, in an accelerator, we have to deal with many dependencies and subsystems. This paper is a result of several years investigations ending up in only few months of dedicated experimentation. In fact, since the discovery of the problem in 2017, the prime purpose was to find a quick, easy and efficient solution, to approve it and deploy it. This has been done with success and allowed the commissioning of the LINAC and its current ramp up. However, we know now that our operation conditions can change (thermal load, pressure, liquid helium level) and that for some operation condition, the TAO wave can appear again and cause pressure instabilities. We also know that the chosen solution is not perfect because of the liquid helium level observed plateau. After deploying the vapour sky short circuit line, our priority was to be able to monitor this phenomena that remained unseen for years, hence the deployment of a simultaneous acquisition of piezo-electric pressure sensors at different locations. The acquisition system is now evolving from the NI-DAQ solution to an integrated programmable ARM DAQ (MSX-E3601) compatible with ModBus TCP communication protocol for a continuous operation compatible with the accelerator data acquisition and archiving system. The analysis pipeline has also to be included in the main data analysis process in order to have alarm kind surveillance of occurring events. A first analysis using Rott's approach and wave approach showed that the system is complex and that a different method has to be considered if we are to understand and find a better solution to the problem. We therefore plan on starting an R\&D program for a dedicated code for analysing cryogenic thermo-acoustics in complex geometries. At the same-time, we plan on deploying a tuneable, damper like test setup that mimics variable resistance, inductance and capacitance thermodynamic RLC circuit in order to better characterise the impedance of a TAO loaded cryogenic system.

\subsection*{Acknowledgements}
This work has been funded by "Region Normandie" as well as the city of Caen, CNRS and CEA. We would like to thank all contributors from CEA-IRFU, CNRS-JC Lab and GANIL without whom this paper would not have been possible. We also thank F. Bonne and P. Bonnay (DSBT/CEA) for the Simcryogenics library that is being used for the model base control of the cryogenics. We thank F. Millet (DSBT/CEA) for useful discussions on liquid helium level sensors shielding. We finally thank D. Longuevergne and M. Pierens from IJC Lab for kindly borrowing us the first fast sensors and acquisition system for vibrations investigations in 2017 and useful discussions on setting up the first experiment.

\bibliography{ref}

\begin{thebibliography}{23}%
\makeatletter
\providecommand \@ifxundefined [1]{%
 \@ifx{#1\undefined}
}%
\providecommand \@ifnum [1]{%
 \ifnum #1\expandafter \@firstoftwo
 \else \expandafter \@secondoftwo
 \fi
}%
\providecommand \@ifx [1]{%
 \ifx #1\expandafter \@firstoftwo
 \else \expandafter \@secondoftwo
 \fi
}%
\providecommand \natexlab [1]{#1}%
\providecommand \enquote  [1]{``#1''}%
\providecommand \bibnamefont  [1]{#1}%
\providecommand \bibfnamefont [1]{#1}%
\providecommand \citenamefont [1]{#1}%
\providecommand \href@noop [0]{\@secondoftwo}%
\providecommand \href [0]{\begingroup \@sanitize@url \@href}%
\providecommand \@href[1]{\@@startlink{#1}\@@href}%
\providecommand \@@href[1]{\endgroup#1\@@endlink}%
\providecommand \@sanitize@url [0]{\catcode `\\12\catcode `\$12\catcode
  `\&12\catcode `\#12\catcode `\^12\catcode `\_12\catcode `\%12\relax}%
\providecommand \@@startlink[1]{}%
\providecommand \@@endlink[0]{}%
\providecommand \url  [0]{\begingroup\@sanitize@url \@url }%
\providecommand \@url [1]{\endgroup\@href {#1}{\urlprefix }}%
\providecommand \urlprefix  [0]{URL }%
\providecommand \Eprint [0]{\href }%
\providecommand \doibase [0]{http://dx.doi.org/}%
\providecommand \selectlanguage [0]{\@gobble}%
\providecommand \bibinfo  [0]{\@secondoftwo}%
\providecommand \bibfield  [0]{\@secondoftwo}%
\providecommand \translation [1]{[#1]}%
\providecommand \BibitemOpen [0]{}%
\providecommand \bibitemStop [0]{}%
\providecommand \bibitemNoStop [0]{.\EOS\space}%
\providecommand \EOS [0]{\spacefactor3000\relax}%
\providecommand \BibitemShut  [1]{\csname bibitem#1\endcsname}%
\let\auto@bib@innerbib\@empty
\bibitem [{\citenamefont {Lewitowicz}(2006)}]{Lewitowicz:2006fx}%
  \BibitemOpen
  \bibfield  {author} {\bibinfo {author} {\bibfnamefont {M.}~\bibnamefont
  {Lewitowicz}},\ }\href@noop {} {\bibfield  {journal} {\bibinfo  {journal}
  {TOURS SYMPOSIUM ON NUCLEAR PHYSICS VI}\ }\textbf {\bibinfo {volume} {891}},\
  \bibinfo {pages} {91} (\bibinfo {year} {2006})}\BibitemShut {NoStop}%
\bibitem [{\citenamefont {Ferdinand}\ \emph {et~al.}(2013)\citenamefont
  {Ferdinand}, \citenamefont {Bernaudin}, \citenamefont {Di~Giacomo},
  \citenamefont {Bosland}, \citenamefont {Olry},\ and\ \citenamefont
  {Gomez~Martinez}}]{ferdinand:in2p3-00867502}%
  \BibitemOpen
  \bibfield  {author} {\bibinfo {author} {\bibfnamefont {R.}~\bibnamefont
  {Ferdinand}}, \bibinfo {author} {\bibfnamefont {P.}~\bibnamefont
  {Bernaudin}}, \bibinfo {author} {\bibfnamefont {M.}~\bibnamefont
  {Di~Giacomo}}, \bibinfo {author} {\bibfnamefont {P.}~\bibnamefont {Bosland}},
  \bibinfo {author} {\bibfnamefont {G.}~\bibnamefont {Olry}}, \ and\ \bibinfo
  {author} {\bibfnamefont {Y.}~\bibnamefont {Gomez~Martinez}},\ }in\ \href
  {http://hal.in2p3.fr/in2p3-00867502} {\emph {\bibinfo {booktitle} {{16th
  International Conference on RF Superconductivity (SRF2013)}}}},\ Vol.\
  \bibinfo {volume} {SRF13}\ (\bibinfo  {publisher} {{Joint Accelerator
  Conferences Website}},\ \bibinfo {address} {Paris, France},\ \bibinfo {year}
  {2013})\ pp.\ \bibinfo {pages} {11--17},\ \bibinfo {note}
  {mOIOA02}\BibitemShut {NoStop}%
\bibitem [{\citenamefont {Dolegieviez}\ \emph {et~al.}(2019)\citenamefont
  {Dolegieviez}, \citenamefont {Ferdinand}, \citenamefont {Ledoux},
  \citenamefont {Savajols},\ and\ \citenamefont
  {Varenne}}]{dolegieviez:hal-02187926}%
  \BibitemOpen
  \bibfield  {author} {\bibinfo {author} {\bibfnamefont {P.}~\bibnamefont
  {Dolegieviez}}, \bibinfo {author} {\bibfnamefont {R.}~\bibnamefont
  {Ferdinand}}, \bibinfo {author} {\bibfnamefont {X.}~\bibnamefont {Ledoux}},
  \bibinfo {author} {\bibfnamefont {H.}~\bibnamefont {Savajols}}, \ and\
  \bibinfo {author} {\bibfnamefont {F.}~\bibnamefont {Varenne}},\ }in\ \href
  {\doibase 10.18429/JACoW-IPAC2019-MOPTS005} {\emph {\bibinfo {booktitle}
  {{10th International Particle Accelerator Conference}}}}\ (\bibinfo {address}
  {Melbourne, Australia},\ \bibinfo {year} {2019})\ p.\ \bibinfo {pages}
  {MOPTS005}\BibitemShut {NoStop}%
\bibitem [{\citenamefont {Ghribi}\ \emph
  {et~al.}(2017{\natexlab{a}})\citenamefont {Ghribi}, \citenamefont
  {Bernaudin}, \citenamefont {Vassal},\ and\ \citenamefont
  {Bonne}}]{ghribi2017}%
  \BibitemOpen
  \bibfield  {author} {\bibinfo {author} {\bibfnamefont {A.}~\bibnamefont
  {Ghribi}}, \bibinfo {author} {\bibfnamefont {P.~E.}\ \bibnamefont
  {Bernaudin}}, \bibinfo {author} {\bibfnamefont {A.}~\bibnamefont {Vassal}}, \
  and\ \bibinfo {author} {\bibfnamefont {F.}~\bibnamefont {Bonne}},\
  }\href@noop {} {\bibfield  {journal} {\bibinfo  {journal} {Cryogenics}\
  }\textbf {\bibinfo {volume} {85}},\ \bibinfo {pages} {44} (\bibinfo {year}
  {2017}{\natexlab{a}})}\BibitemShut {NoStop}%
\bibitem [{\citenamefont {Ghribi}\ \emph
  {et~al.}(2017{\natexlab{b}})\citenamefont {Ghribi}, \citenamefont
  {Bernaudin}, \citenamefont {Bert}, \citenamefont {Commeaux}, \citenamefont
  {Houeto},\ and\ \citenamefont {Lescalié}}]{ghribi2017_2}%
  \BibitemOpen
  \bibfield  {author} {\bibinfo {author} {\bibfnamefont {A.}~\bibnamefont
  {Ghribi}}, \bibinfo {author} {\bibfnamefont {P.-E.}\ \bibnamefont
  {Bernaudin}}, \bibinfo {author} {\bibfnamefont {Y.}~\bibnamefont {Bert}},
  \bibinfo {author} {\bibfnamefont {C.}~\bibnamefont {Commeaux}}, \bibinfo
  {author} {\bibfnamefont {M.}~\bibnamefont {Houeto}}, \ and\ \bibinfo {author}
  {\bibfnamefont {G.}~\bibnamefont {Lescalié}},\ }\href@noop {} {\bibfield
  {journal} {\bibinfo  {journal} {IOP Conference Series: Materials Science and
  Engineering}\ }\textbf {\bibinfo {volume} {171}},\ \bibinfo {pages} {012115}
  (\bibinfo {year} {2017}{\natexlab{b}})}\BibitemShut {NoStop}%
\bibitem [{\citenamefont {Vassal}\ \emph {et~al.}(2019)\citenamefont {Vassal},
  \citenamefont {Bonne}, \citenamefont {Ghribi}, \citenamefont {Millet},
  \citenamefont {Bonnay},\ and\ \citenamefont {Bernaudin}}]{Vassal_2019}%
  \BibitemOpen
  \bibfield  {author} {\bibinfo {author} {\bibfnamefont {A.}~\bibnamefont
  {Vassal}}, \bibinfo {author} {\bibfnamefont {F.}~\bibnamefont {Bonne}},
  \bibinfo {author} {\bibfnamefont {A.}~\bibnamefont {Ghribi}}, \bibinfo
  {author} {\bibfnamefont {F.}~\bibnamefont {Millet}}, \bibinfo {author}
  {\bibfnamefont {P.}~\bibnamefont {Bonnay}}, \ and\ \bibinfo {author}
  {\bibfnamefont {P.-E.}\ \bibnamefont {Bernaudin}},\ }\href {\doibase
  10.1088/1757-899x/502/1/012111} {\bibfield  {journal} {\bibinfo  {journal}
  {{IOP} Conference Series: Materials Science and Engineering}\ }\textbf
  {\bibinfo {volume} {502}},\ \bibinfo {pages} {012111} (\bibinfo {year}
  {2019})}\BibitemShut {NoStop}%
\bibitem [{\citenamefont {Bonne}\ \emph {et~al.}(2020)\citenamefont {Bonne},
  \citenamefont {Varin}, \citenamefont {Vassal}, \citenamefont {Bonnay},
  \citenamefont {Hoa}, \citenamefont {Millet},\ and\ \citenamefont
  {Poncet}}]{Bonne_2020}%
  \BibitemOpen
  \bibfield  {author} {\bibinfo {author} {\bibfnamefont {F.}~\bibnamefont
  {Bonne}}, \bibinfo {author} {\bibfnamefont {S.}~\bibnamefont {Varin}},
  \bibinfo {author} {\bibfnamefont {A.}~\bibnamefont {Vassal}}, \bibinfo
  {author} {\bibfnamefont {P.}~\bibnamefont {Bonnay}}, \bibinfo {author}
  {\bibfnamefont {C.}~\bibnamefont {Hoa}}, \bibinfo {author} {\bibfnamefont
  {F.}~\bibnamefont {Millet}}, \ and\ \bibinfo {author} {\bibfnamefont {J.-M.}\
  \bibnamefont {Poncet}},\ }\href {\doibase 10.1088/1757-899x/755/1/012076}
  {\bibfield  {journal} {\bibinfo  {journal} {{IOP} Conference Series:
  Materials Science and Engineering}\ }\textbf {\bibinfo {volume} {755}},\
  \bibinfo {pages} {012076} (\bibinfo {year} {2020})}\BibitemShut {NoStop}%
\bibitem [{\citenamefont {Taconis}\ \emph {et~al.}(1949)\citenamefont
  {Taconis}, \citenamefont {Beenakker}, \citenamefont {Nier},\ and\
  \citenamefont {Aldrich}}]{TACONIS1949733}%
  \BibitemOpen
  \bibfield  {author} {\bibinfo {author} {\bibfnamefont {K.}~\bibnamefont
  {Taconis}}, \bibinfo {author} {\bibfnamefont {J.}~\bibnamefont {Beenakker}},
  \bibinfo {author} {\bibfnamefont {A.}~\bibnamefont {Nier}}, \ and\ \bibinfo
  {author} {\bibfnamefont {L.}~\bibnamefont {Aldrich}},\ }\href {\doibase
  https://doi.org/10.1016/0031-8914(49)90078-6} {\bibfield  {journal} {\bibinfo
   {journal} {Physica}\ }\textbf {\bibinfo {volume} {15}},\ \bibinfo {pages}
  {733 } (\bibinfo {year} {1949})}\BibitemShut {NoStop}%
\bibitem [{\citenamefont {Swift}(2007)}]{Swift2007}%
  \BibitemOpen
  \bibfield  {author} {\bibinfo {author} {\bibfnamefont {G.}~\bibnamefont
  {Swift}},\ }\enquote {\bibinfo {title} {Thermoacoustics},}\ in\ \href
  {\doibase 10.1007/978-0-387-30425-0_7} {\emph {\bibinfo {booktitle} {Springer
  Handbook of Acoustics}}},\ \bibinfo {editor} {edited by\ \bibinfo {editor}
  {\bibfnamefont {T.}~\bibnamefont {Rossing}}}\ (\bibinfo  {publisher}
  {Springer New York},\ \bibinfo {address} {New York, NY},\ \bibinfo {year}
  {2007})\ pp.\ \bibinfo {pages} {239--255}\BibitemShut {NoStop}%
\bibitem [{\citenamefont {A.}(1949)}]{kramers1949}%
  \BibitemOpen
  \bibfield  {author} {\bibinfo {author} {\bibfnamefont {K.~H.}\ \bibnamefont
  {A.}},\ }\href@noop {} {\bibfield  {journal} {\bibinfo  {journal} {Physica
  (Amsterdam)}\ }\textbf {\bibinfo {volume} {15}},\ \bibinfo {pages} {971}
  (\bibinfo {year} {1949})}\BibitemShut {NoStop}%
\bibitem [{\citenamefont {N.}(1969)}]{rott1969}%
  \BibitemOpen
  \bibfield  {author} {\bibinfo {author} {\bibfnamefont {R.}~\bibnamefont
  {N.}},\ }\href@noop {} {\bibfield  {journal} {\bibinfo  {journal} {Z. Angew.
  Math. Phys.}\ }\textbf {\bibinfo {volume} {20}},\ \bibinfo {pages} {230}
  (\bibinfo {year} {1969})}\BibitemShut {NoStop}%
\bibitem [{\citenamefont {N.}(1973)}]{rott1973}%
  \BibitemOpen
  \bibfield  {author} {\bibinfo {author} {\bibfnamefont {R.}~\bibnamefont
  {N.}},\ }\href@noop {} {\bibfield  {journal} {\bibinfo  {journal} {Z. Angew.
  Math. Phys.}\ }\textbf {\bibinfo {volume} {24}},\ \bibinfo {pages} {54}
  (\bibinfo {year} {1973})}\BibitemShut {NoStop}%
\bibitem [{\citenamefont {Sugimoto}\ and\ \citenamefont
  {Shimizu}(2008)}]{doi:10.1063/1.2990763}%
  \BibitemOpen
  \bibfield  {author} {\bibinfo {author} {\bibfnamefont {N.}~\bibnamefont
  {Sugimoto}}\ and\ \bibinfo {author} {\bibfnamefont {D.}~\bibnamefont
  {Shimizu}},\ }\href {\doibase 10.1063/1.2990763} {\bibfield  {journal}
  {\bibinfo  {journal} {Physics of Fluids}\ }\textbf {\bibinfo {volume} {20}},\
  \bibinfo {pages} {104102} (\bibinfo {year} {2008})},\ \Eprint
  {http://arxiv.org/abs/https://doi.org/10.1063/1.2990763}
  {https://doi.org/10.1063/1.2990763} \BibitemShut {NoStop}%
\bibitem [{\citenamefont {Sun}\ \emph {et~al.}(2016)\citenamefont {Sun},
  \citenamefont {Wang}, \citenamefont {Guo}, \citenamefont {Zhang},
  \citenamefont {Xu}, \citenamefont {Zou},\ and\ \citenamefont
  {Zhang}}]{SUN201638}%
  \BibitemOpen
  \bibfield  {author} {\bibinfo {author} {\bibfnamefont {D.}~\bibnamefont
  {Sun}}, \bibinfo {author} {\bibfnamefont {K.}~\bibnamefont {Wang}}, \bibinfo
  {author} {\bibfnamefont {Y.}~\bibnamefont {Guo}}, \bibinfo {author}
  {\bibfnamefont {J.}~\bibnamefont {Zhang}}, \bibinfo {author} {\bibfnamefont
  {Y.}~\bibnamefont {Xu}}, \bibinfo {author} {\bibfnamefont {J.}~\bibnamefont
  {Zou}}, \ and\ \bibinfo {author} {\bibfnamefont {X.}~\bibnamefont {Zhang}},\
  }\href {\doibase https://doi.org/10.1016/j.cryogenics.2016.01.004} {\bibfield
   {journal} {\bibinfo  {journal} {Cryogenics}\ }\textbf {\bibinfo {volume}
  {75}},\ \bibinfo {pages} {38 } (\bibinfo {year} {2016})}\BibitemShut
  {NoStop}%
\bibitem [{\citenamefont {Ghribi}\ \emph
  {et~al.}(2017{\natexlab{c}})\citenamefont {Ghribi}, \citenamefont
  {Bernaudin}, \citenamefont {Berthe}, \citenamefont {Duteil}, \citenamefont
  {Ferdinand}, \citenamefont {Lescali{\'e}}, \citenamefont {Philippe},
  \citenamefont {Thivel}, \citenamefont {Trudel}, \citenamefont {Vassal},
  \citenamefont {Roz{\'e}},\ and\ \citenamefont
  {Touchard}}]{ghribi:in2p3-01569768}%
  \BibitemOpen
  \bibfield  {author} {\bibinfo {author} {\bibfnamefont {A.}~\bibnamefont
  {Ghribi}}, \bibinfo {author} {\bibfnamefont {P.}~\bibnamefont {Bernaudin}},
  \bibinfo {author} {\bibfnamefont {C.}~\bibnamefont {Berthe}}, \bibinfo
  {author} {\bibfnamefont {G.}~\bibnamefont {Duteil}}, \bibinfo {author}
  {\bibfnamefont {R.}~\bibnamefont {Ferdinand}}, \bibinfo {author}
  {\bibfnamefont {G.}~\bibnamefont {Lescali{\'e}}}, \bibinfo {author}
  {\bibfnamefont {L.}~\bibnamefont {Philippe}}, \bibinfo {author}
  {\bibfnamefont {Y.}~\bibnamefont {Thivel}}, \bibinfo {author} {\bibfnamefont
  {A.}~\bibnamefont {Trudel}}, \bibinfo {author} {\bibfnamefont
  {A.}~\bibnamefont {Vassal}}, \bibinfo {author} {\bibfnamefont
  {J.}~\bibnamefont {Roz{\'e}}}, \ and\ \bibinfo {author} {\bibfnamefont
  {D.}~\bibnamefont {Touchard}},\ }\href {http://hal.in2p3.fr/in2p3-01569768}
  {\  (\bibinfo {year} {2017}{\natexlab{c}})},\ \bibinfo {note}
  {poster}\BibitemShut {NoStop}%
\bibitem [{\citenamefont {Ghribi}\ \emph {et~al.}(2020)\citenamefont {Ghribi},
  \citenamefont {Aburas}, \citenamefont {Baumont}, \citenamefont {Bernaudin},
  \citenamefont {Bonneau}, \citenamefont {Duteil}, \citenamefont {Ferdinand},
  \citenamefont {Lechartier}, \citenamefont {Leyge}, \citenamefont {Lescalié},
  \citenamefont {Thivel}, \citenamefont {Trudel}, \citenamefont {Valentin},\
  and\ \citenamefont {Vassal}}]{GHRIBI2020103126}%
  \BibitemOpen
  \bibfield  {author} {\bibinfo {author} {\bibfnamefont {A.}~\bibnamefont
  {Ghribi}}, \bibinfo {author} {\bibfnamefont {M.}~\bibnamefont {Aburas}},
  \bibinfo {author} {\bibfnamefont {Y.}~\bibnamefont {Baumont}}, \bibinfo
  {author} {\bibfnamefont {P.-E.}\ \bibnamefont {Bernaudin}}, \bibinfo {author}
  {\bibfnamefont {S.}~\bibnamefont {Bonneau}}, \bibinfo {author} {\bibfnamefont
  {G.}~\bibnamefont {Duteil}}, \bibinfo {author} {\bibfnamefont
  {R.}~\bibnamefont {Ferdinand}}, \bibinfo {author} {\bibfnamefont
  {M.}~\bibnamefont {Lechartier}}, \bibinfo {author} {\bibfnamefont {J.-F.}\
  \bibnamefont {Leyge}}, \bibinfo {author} {\bibfnamefont {G.}~\bibnamefont
  {Lescalié}}, \bibinfo {author} {\bibfnamefont {Y.}~\bibnamefont {Thivel}},
  \bibinfo {author} {\bibfnamefont {A.}~\bibnamefont {Trudel}}, \bibinfo
  {author} {\bibfnamefont {L.}~\bibnamefont {Valentin}}, \ and\ \bibinfo
  {author} {\bibfnamefont {A.}~\bibnamefont {Vassal}},\ }\href {\doibase
  https://doi.org/10.1016/j.cryogenics.2020.103126} {\bibfield  {journal}
  {\bibinfo  {journal} {Cryogenics}\ }\textbf {\bibinfo {volume} {110}},\
  \bibinfo {pages} {103126} (\bibinfo {year} {2020})}\BibitemShut {NoStop}%
\bibitem [{\citenamefont {Gu}\ and\ \citenamefont
  {Timmerhaus}(1992)}]{GU1992194}%
  \BibitemOpen
  \bibfield  {author} {\bibinfo {author} {\bibfnamefont {Y.}~\bibnamefont
  {Gu}}\ and\ \bibinfo {author} {\bibfnamefont {K.}~\bibnamefont
  {Timmerhaus}},\ }\href {\doibase
  https://doi.org/10.1016/0011-2275(92)90266-D} {\bibfield  {journal} {\bibinfo
   {journal} {Cryogenics}\ }\textbf {\bibinfo {volume} {32}},\ \bibinfo {pages}
  {194 } (\bibinfo {year} {1992})},\ \bibinfo {note} {space Cryogenics
  Workshop}\BibitemShut {NoStop}%
\bibitem [{\citenamefont {Ditmars}(1965)}]{ditmars1965}%
  \BibitemOpen
  \bibfield  {author} {\bibinfo {author} {\bibfnamefont {G.}~\bibnamefont
  {Ditmars}, \bibfnamefont {D.A.;~Furukawa}},\ }\href@noop {} {\bibfield
  {journal} {\bibinfo  {journal} {Journal of Research of the National Bureau of
  Standards}\ }\textbf {\bibinfo {volume} {69C}},\ \bibinfo {pages} {35}
  (\bibinfo {year} {1965})}\BibitemShut {NoStop}%
\bibitem [{\citenamefont {Chen}\ and\ \citenamefont {Jin}(1999)}]{CHEN1999843}%
  \BibitemOpen
  \bibfield  {author} {\bibinfo {author} {\bibfnamefont {G.}~\bibnamefont
  {Chen}}\ and\ \bibinfo {author} {\bibfnamefont {T.}~\bibnamefont {Jin}},\
  }\href {\doibase https://doi.org/10.1016/S0011-2275(99)00099-5} {\bibfield
  {journal} {\bibinfo  {journal} {Cryogenics}\ }\textbf {\bibinfo {volume}
  {39}},\ \bibinfo {pages} {843 } (\bibinfo {year} {1999})}\BibitemShut
  {NoStop}%
\bibitem [{\citenamefont {Luck}\ and\ \citenamefont
  {Trepp}(1992)}]{LUCK1992703}%
  \BibitemOpen
  \bibfield  {author} {\bibinfo {author} {\bibfnamefont {H.}~\bibnamefont
  {Luck}}\ and\ \bibinfo {author} {\bibfnamefont {C.}~\bibnamefont {Trepp}},\
  }\href {\doibase https://doi.org/10.1016/0011-2275(92)90279-J} {\bibfield
  {journal} {\bibinfo  {journal} {Cryogenics}\ }\textbf {\bibinfo {volume}
  {32}},\ \bibinfo {pages} {703 } (\bibinfo {year} {1992})}\BibitemShut
  {NoStop}%
\bibitem [{\citenamefont {Rott}(1980)}]{ROTT1980135}%
  \BibitemOpen
  \bibfield  {author} {\bibinfo {author} {\bibfnamefont {N.}~\bibnamefont
  {Rott}},\ }\href {\doibase https://doi.org/10.1016/S0065-2156(08)70233-3}
  {\emph {\bibinfo {title} {Thermoacoustics}}},\ edited by\ \bibinfo {editor}
  {\bibfnamefont {C.-S.}\ \bibnamefont {Yih}},\ \bibinfo {series} {Advances in
  Applied Mechanics}, Vol.~\bibinfo {volume} {20}\ (\bibinfo  {publisher}
  {Elsevier},\ \bibinfo {year} {1980})\ pp.\ \bibinfo {pages} {135 --
  175}\BibitemShut {NoStop}%
\bibitem [{\citenamefont {Gupta}\ and\ \citenamefont
  {Rabehl}(2015)}]{GUPTA2015104}%
  \BibitemOpen
  \bibfield  {author} {\bibinfo {author} {\bibfnamefont {P.~K.}\ \bibnamefont
  {Gupta}}\ and\ \bibinfo {author} {\bibfnamefont {R.}~\bibnamefont {Rabehl}},\
  }\href {\doibase https://doi.org/10.1016/j.applthermaleng.2015.03.051}
  {\bibfield  {journal} {\bibinfo  {journal} {Applied Thermal Engineering}\
  }\textbf {\bibinfo {volume} {84}},\ \bibinfo {pages} {104 } (\bibinfo {year}
  {2015})}\BibitemShut {NoStop}%
\bibitem [{\citenamefont {J.W.~Strutt}(1877)}]{KU2005588}%
  \BibitemOpen
  \bibfield  {author} {\bibinfo {author} {\bibfnamefont {t.~B.~R.}\
  \bibnamefont {J.W.~Strutt}},\ }\href@noop {} {\emph {\bibinfo {title} {The
  Theory of Sound}}},\ edited by\ \bibinfo {editor} {\bibnamefont {Macmillan}}\
  and\ \bibinfo {editor} {\bibnamefont {co.}}\ (\bibinfo  {publisher}
  {Cambridge University Press},\ \bibinfo {year} {1877})\BibitemShut {NoStop}%
\end{thebibliography}%

\end{document}